\documentclass[reprint,amsmath,amssymb,
 aps,prb]{revtex4-2}

\usepackage{graphicx}
\graphicspath{{figures/}}
\usepackage{dcolumn}
\usepackage{bm}
\usepackage{hyperref}
\usepackage{dsfont}
\usepackage{wrapfig}
\usepackage{physics}
\usepackage{xcolor}

\usepackage{subcaption}
\usepackage{cancel}
\usepackage{verbatim}
\usepackage{dsfont}
\newcommand{\unit}[1]{\ensuremath{\, \mathrm{#1}}}
\begin{document}

\title{Driven Andreev molecule}

\author{Andriani Keliri}\email{akeliri@lpthe.jussieu.fr}
 \affiliation{Laboratoire de Physique Th\'{e}orique et Hautes Energies,
Sorbonne Universit\'{e} and CNRS UMR 7589, 4 place Jussieu, 75252 Paris Cedex 05, France}
\author{Beno\^{i}t Dou\c{c}ot}
 \affiliation{Laboratoire de Physique Th\'{e}orique et Hautes Energies,
Sorbonne Universit\'{e} and CNRS UMR 7589, 4 place Jussieu, 75252 Paris Cedex 05, France}

\date{\today}

\begin{abstract}
We study the three terminal S-QD-S-QD-S Josephson junction biased with commensurate voltages. In the absence of an applied voltage, the Andreev bound states on each quantum dot hybridize forming an `Andreev molecule’. However, understanding of this system in a non-equilibrium setup is lacking. Applying a dc voltage on the bijunction makes the system time-periodic, and the equilibrium Andreev bound states evolve into a ladder of resonances with a finite lifetime due to multiple Andreev reflections (MAR). Starting from the time-periodic Bogoliubov-de Gennes equations we map the problem to a tight-binding chain in the (infinite) Floquet space. The resolvent of this non-Hermitian block matrix is obtained via a continued fraction method. We numerically calculate the Floquet-Andreev spectra which could be probed by local tunneling spectroscopy on the dots. We also consider the subgap current, and show that the Floquet resonances determine the position of the MAR steps. Proximity of the two dots causes splitting of the steps, while at large distances we observe interference effects which cause oscillations in the I--V curves. The latter effect should persist at very long distances.
\end{abstract}

\maketitle

\section{Introduction}
At its heart, superconductivity has the distinct characteristic of a coherent macroscopic quantum state. Together with that other particularly quantum phenomenon -- magnetism -- superconductivity is bound to be an essential element for quantum computing applications \cite{devoret}. The working principle of such superconducting circuits is the Josephson effect(s) \cite{Josephson,josephson_observation}. A more microscopic description of Josephson junctions reveals the Andreev reflection mechanism and the resulting Andreev bound states (ABS) as responsible for carrying most of the Josephson current across a phase-biased junction. Apart from using the ABS to realize an `Andreev qubit' \cite{andreevqubit,coherent}, other, more complex geometries are being explored. Multi-terminal Josephson junctions are drawing particular interest since, on the one hand, they could be an alternative way to engineer topological states, even when the junctions are made from topologically trivial materials \cite{riwar2016multi,berry}, and on the other, they can be used to create correlations among pairs of Cooper pairs, the so-called quartets \cite{quartets1,quartets2,quartets3,quartets4}.\par
In an analogy to the formation of a molecule, bringing two ABS carrying junctions close enough should result in a hybridization of the ABS wave-functions \cite{molecule, scattering}. The hybridization would create non-local effects in the current, whereby changing the phase on one junction would change the current flowing through the other. This could be useful for realizing qubits whose coupling, for example, can be tuned by changing their phase difference, but one should have a distance of the junctions which remains comparable to the superconducting coherence length $\xi_0.$ In a typical superconductor, such as aluminum, $\xi_0\sim 100 \unit{nm}.$ The first experiments realizing Andreev molecules and measuring non-local effects in the Josephson current have already been performed on semiconducting nanowires \cite{InAs1,InAs2}. However, such systems host subgap states which are different from the ABS (the Yu-Shiba-Rusinov states) because they are in the opposite limit of strong Coulomb interaction ($U>\Delta$). Another interesting proposal is to use an Andreev molecule as an elementary unit for realizing a physical Kitaev chain, which could host the elusive Majorana state \cite{kitaevchain,poormajorana}.\par
Meanwhile, there is increasing interest in periodically driven (Floquet) systems since the external drive can be used to engineer new `hidden' states and dynamically control properties otherwise inaccessible in equilibrium \cite{kohler,floquetengineering,topologicalinsulators,rudner}. For example, to open band-gaps in graphene \cite{graphene1}, or induce edge-states that carry an anomalous Hall current \cite{graphene2}. Moreover, Floquet qubits would offer numerous optimal working points to choose from by changing the driving parameters \cite{sweetspots, fluxonium, floquetqubit}, contrary to their static counterparts whose parameters are mostly tuned during fabrication. Experimentally, the realization of Floquet states is often difficult due to thermalization \cite{lazarides} and short lifetimes. Nevertheless, a recent experiment has reported the generation of long-lived steady Floquet–Andreev states realized by continuous microwave irradiation of a graphene Josephson junction \cite{steady}.\par
Periodic driving of a superconducting junction could also be realized by voltage-biasing the junction. It has been known since the 80's that, in such cases, multiple Andreev reflections (MAR) of quasi-particles between the junctions' superconductors lead to a sub-harmonic gap structure of the current-voltage characteristics: the current exhibits jumps at particular voltage values which are integer subdivisions of the superconducting gap $eV=2\Delta/n$ \cite{BTK,subgap}. In the limit of resonant tunneling through the junction the subgap structure is greatly modified \cite{yeyati-dot,shumeiko,Jonckheere,QDreview}: features corresponding to an odd number of MAR are enhanced, while even trajectories are suppressed. Previous work has shown that, in this case, the equilibrium ABS evolve into ladders of resonances \cite{fwsa,engineering} equivalent to the Wannier-Stark ladders of a solid in an electric field \cite{wannier,wannier-stark}.\par
In this work, we study the Floquet spectra of a driven Andreev molecule, in the limit of resonant tunneling through the junctions. We therefore model each junction by a quantum dot. Using a quantum scattering theory approach, we map the problem to a tight-binding chain in the infinite Floquet space. We see that the equilibrium ABS on each dot evolve into ladders of resonances exhibiting level splitting in the molecular regime (when the separation between the dots is comparable to the superconducting coherence length, $R\sim \xi_0$). These Floquet-Andreev resonances should leave their trace in the dc current. We therefore calculate the steady-state current passing through one junction and see that the proximity of the second junction modifies the usual MAR steps, which accordingly exhibit splitting into sub-steps. Moreover, when the two junctions are separated by a large distance $R\gg\xi_0$, we find oscillations of the spectral functions above the gap and, consequently, of the I-V curves. This phenomenon is akin to the Tomasch effect \cite{Tomasch,McMillan-Anderson}, and is due to long-distance correlations mediated by propagating quasi-particles in the middle superconductor, as has been discussed in recent work \cite{melin2021}. Indeed, we find that the spectral function at a fixed voltage value is an oscillatory function of a Tomasch phase factor. This Floquet-Tomasch effect should persist at distances which are up to two orders of magnitude larger than the superconducting coherence length, as in the Tomasch experiment.\par
The rest of this paper has the following structure. In Sec. \ref{model} we define the model Hamiltonian and derive the Floquet-Lippmann-Scwinger equations. We then show how to calculate the resolvent operator and the subgap current. Section \ref{spectroscopy} presents our numerical results for the Floquet spectra. Section \ref{current} shows our results for the subgap current, first for the current through a single junction, and then for the Andreev molecule. Conclusions and perspectives are provided in Sect. \ref{conclusions}. We show that the spectral function oscillates due to a Tomasch phase factor in Appendix \ref{oscillations}.

\begin{figure}[b]
    \includegraphics[width=\linewidth]{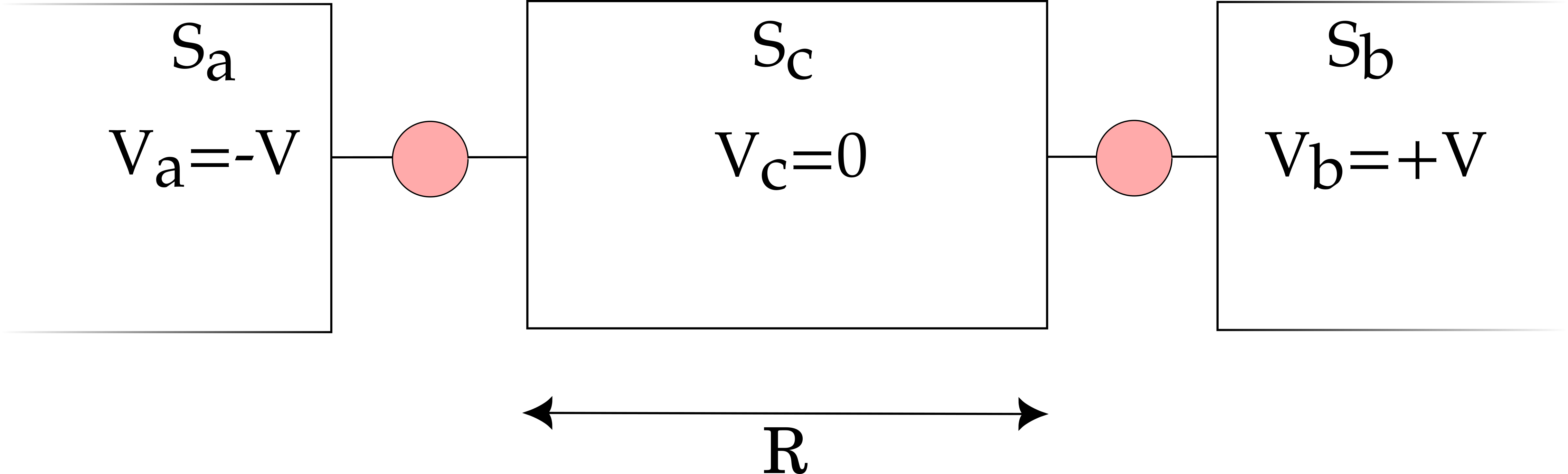}    
    \caption{Schematic representation of the three-terminal Josephson junction setup considered in this paper. The two quantum dots host a single discrete level at zero energy and are separated through the middle superconductor $S_c$ by a distance $R.$ The reservoirs are considered to be one-dimensional and are voltage-biased along the `quartet line': $(V_a,V_c,V_b)=(-V,0,+V).$}
    \label{sketch}
\end{figure}

\section{Model and method}\label{model}

\subsection{Hamiltonian}
The system considered here is the three-terminal Josephson junction, with two quantum dots each connected to a superconducting reservoir $S_a, S_b,$ and both connected to a central superconductor $S_c,$ as illustrated in Fig. \ref{sketch}. We study the simplest case where each dot is modeled by a single discrete resonant level at energy $\epsilon_d=0.$ When a dc voltage $V$ is applied across a superconducting junction its Hamiltonian acquires a time-periodic dependence according to the Josephson relation $\phi(t)=\phi(0)+\frac{2eV}{\hbar}t,$ where $\phi$ is the superconducting phase difference across the junction. In this paper, we use the so-called quartet configuration when biasing the superconducting reservoirs: $(V_a,V_c,V_b)=(-V,0,+V).$ This choice makes the problem simpler since it leads to a single basic frequency $\omega_0=eV/\hbar$ for the whole system. The resulting Hamiltonian has the discrete symmetry $\mathcal{H}(t)=\mathcal{H}(t+T),$ where $T=\frac{2\pi}{\omega_0}$ is the period of the drive. A simple gauge transformation then permits to write the Hamiltonian in the form:
\begin{equation}
    \mathcal{H}(t)= \mathcal{H}_0 +\mathcal{V}(t).
\end{equation} 
The static part $\mathcal{H}_0$ is a sum of the BCS Hamiltonians describing the superconducting reservoirs: 
\begin{equation}\label{bcs}
\mathcal{H}_0 = \sum_{jk\sigma}\epsilon_{k} c^{\dagger}_{jk\sigma} c_{jk\sigma} +\sum_{jk} \qty(\Delta_j c^{\dagger}_{jk\uparrow} c^{\dagger}_{j-k\downarrow} +\Delta^{\ast}_j c_{j-k\downarrow} c_{jk\uparrow}).
\end{equation}
The operators $c^{\dagger}_{jk\sigma}$ and $c_{jk\sigma}$ create and annihilate an electron in the $j$ reservoir with momentum $k$ and spin $\sigma.$ We use the notation $\Delta_j=\Delta e^{i\phi_j},$ and consider that all superconductors have a gap of equal magnitude. 

The time-periodic part $\mathcal{V}(t)$ describes the tunneling between the dots labeled by $i=\{1,2\}$ and the reservoirs labeled by $j:$
\begin{equation}\label{tunneling}
\begin{split}
 \mathcal{V}(t) =\sum_{i\in\mathrm{dots}}\sum_{jk\sigma} &( J_j(x_i) e^{is_j\omega_0 t} d^{\dagger}_{i\sigma}c_{jk\sigma} \\
 &+ J_j^{\star}(x_i)e^{-is_j\omega_0 t} c^{\dagger}_{jk\sigma}d_{i\sigma})   
\end{split}
\end{equation}
For convenience, we take the dots' positions to be at $x_1=0,x_2=R,$ and the tunnel couplings to be $J_j(x_i)=J_je^{ikx_i},$ with a real amplitude $J_j=J_j^{\star}.$ We have moreover used the notation $V_j=s_jV.$ 

The main idea is to exploit the symmetry of the Hamiltonian by looking for time-periodic (Floquet) solutions to the Bogoliubov-de Gennes (BdG) equations. The problem can then be mapped to an effective tight-binding model \cite{grempel,Dalibard,Holthaus,quantumchaos}, with sites labeled by the Floquet harmonics. In such time-periodic systems the energy is not a well-defined quantity \cite{shirley,zeldovich,sambe}, but is defined modulo the frequency of the drive. We therefore talk of `quasi-energies' of the system, in direct analogy to the quasi-momenta of Bloch theory due to periodicity in space.

\subsection{Floquet-Lippmann-Schwinger equations}
In the absence of tunneling, $V(t)=0,$ the BdG equation can be written as
\begin{equation}
	i\dv{t}\gamma^{\dagger}_{lk\sigma}=\comm{H_0}{\gamma^{\dagger}_{lk\sigma}}=E_{lk}\gamma^{\dagger}_{lk\sigma}.
\end{equation}
The bare quasi-particle operator $\gamma_{lk\sigma}$ is an eigenstate of the superconducting reservoir labeled $l=\qty{a,b,c}:$
\begin{equation}
 \gamma^{\dagger}_{lk\sigma}(t) =e^{-iE_{lk}t}\pqty{ x_{lk} e^{i\phi_l/2} c^{\dagger}_{lk\sigma} +\sigma y_{lk}e^{-i\phi_l/2}c_{l-k-\sigma}},
\end{equation}
and the coefficients $x,y$ are the usual coefficients obtained by diagonalizing the BCS Hamiltonian \cite{coleman}
\begin{equation}
	x^2_{lk}=\frac{E_{lk}+\epsilon_{k}}{2E_{lk}} \qquad \mathrm{and} \qquad
	y^2_{lk}=\frac{E_{lk}-\epsilon_{k}}{2E_{lk}}
\end{equation}
with $E_{lk} \equiv \sqrt{\epsilon^2_k+\abs{\Delta_l}^2}$ the excitation energy needed for adding an electron or a hole to the BCS ground state. 

When the tunneling is turned on, the discrete ABS states on the dots become resonances due to the multiple Andreev reflection processes (MAR) which connect them to the superconducting continua. We can then use the Lippmann-Schwinger method from quantum scattering theory to construct dressed operators $\Gamma$ which tend to the bare quasi-particle operators $\gamma$ at the limit of zero tunnel couplings. We therefore introduce a dressed quasi-particle operator $\Gamma^{\dagger}_{lk\sigma}$ describing a quasi-particle being injected from a source reservoir $l$ with momentum $k$ and spin $\sigma$, to any of the reservoirs $j,$ and quantum dot(s) $i$
\begin{equation}\label{qp operator}
\begin{split}
   &\Gamma^{\dagger}_{lk\sigma}(t) = \gamma^{\dagger}_{lk\sigma}(t)+ e^{-iE_{lk}t} \sum_{m\in\mathbb{Z}} e^{-im\omega_0t} \\
   &\qquad \times\Bigg[ \sum_{i\in\mathrm{dots}}\qty(u_m(i;lk)d^{\dagger}_{i\sigma}+\sigma v_m(i;lk)d_{i-\sigma}) \\
   &+ \sum_{jk'}\qty(U_m(jk';lk)c^{\dagger}_{jk'\sigma}+\sigma V_m(jk';lk)c_{j-k'-\sigma})\Bigg].  
\end{split}
\end{equation}
The amplitudes $u_m(i;lk),v_m(i;lk)$ have respectively the meaning of an electron-- or hole--like amplitude on the dot $i,$ corresponding to a Floquet harmonic with quasi-energy $E_{lk}+m\omega_0,$ while capital letters $U_m,V_m$ correspond to amplitudes in the reservoirs.

The dressed operators are Floquet solutions of the BdG equations
\begin{equation}\label{BdG}
	i\dv{t}\Gamma^{\dagger}_{lk\sigma}(t)=\comm{H(t)}{\Gamma^{\dagger}_{lk\sigma}},
\end{equation}
and therefore obey the Floquet theorem
\begin{equation}
	\Gamma^{\dagger}(t+T)=e^{-iE_{lk}t}\Gamma^{\dagger}(t).
\end{equation}
We use the ansatz of Eq. (\ref{qp operator}) and substitute into Eq. (\ref{BdG}). After integrating out the amplitudes of the reservoirs, we are lead to the following set of in-homogeneous Lippmann-Schwinger equations for the amplitudes on the dots:

\begin{widetext}
\begin{equation}\label{FLS}
	\begin{split}
		(E_{lk}+m\omega_0+i\eta)u_m(i;lk) & - \sum_{ji'}\bqty{\tilde{g}_{j,ii'}^{11}(m+s_j)u_m(i';lk)-\tilde{g}_{j,ii'}^{12}(m+s_j)v_{m+2s_j}(i';lk)} =\delta_{m,-s_l} J_l(x_i) x_{lk}e^{i\phi_l/2}\\
		(E_{lk}+m\omega_0+i\eta)v_m(i;lk) & - \sum_{ji'}\bqty{-\tilde{g}_{j,ii'}^{21}(m-s_j)u_{m-2s_j}(i';lk)+\tilde{g}_{j,ii'}^{22}(m-s_j)v_m(i';lk)} =-\delta_{m,s_l}J_l(x_i) y_{lk}e^{-i\phi_l/2}.
	\end{split}
\end{equation}
\end{widetext}
In the above equation, $\tilde{g}_{j,ii'}\delta_{ii'}\equiv \tilde{g}_j$ is the Green's function for the 1D superconductor of the $j$ reservoir
\begin{equation}\label{greens functions}
	\begin{split}
		\tilde{g}_j(\omega) &= \frac{\Gamma_j}{iv_F q(\omega)} \mqty(\omega & \Delta_j \\ \Delta^{\ast}_j & \omega),\\
		v_F q(\omega) &\equiv i\sqrt{\Delta^2-\omega^2} \theta(\Delta-\abs{\omega}) \\ & \qquad+ \mathrm{sign}(\omega)\sqrt{\omega^2-\Delta^2} \theta(\abs{\omega}-\Delta).
	\end{split}
\end{equation}
We have used the notation $\Gamma_j=\pi\rho_0 J_j^2,$ where $\rho_0$ is the density of states in the normal state of the superconductors. Moreover, since the quasi-energy appears in the combination $\omega+m\omega_0,$ it is convenient to use the shorthand $f(m)$ instead of $f(\omega+m\omega_0),$ for any function $f.$ 
The non-diagonal part which couples the amplitudes on different dots $\tilde{g}_{j,ii'}(m)(1-\delta_{ii'})\equiv \tilde{g}_j(m,R)$ is a non-local Green's function and depends explicitly on the distance between the two dots:
\begin{equation}
	\tilde{g}_j(m,R) = e^{iq(m) R}\qty[\cos(k_F R) \tilde{g}_j(m) +\sin(k_F R) \Gamma_j \sigma_z],
\end{equation}
where $k_F$ is the Fermi wave-vector. 

The non-local Green's function mediates the coupling between the junctions and oscillates on two very different length scales. For energies smaller than the gap $\abs{\omega}<\Delta,$ the factor $e^{iqR}=e^{-\sqrt{1-(\omega/\Delta)^2}R/\xi_0}$ decays exponentially over distances larger than the superconducting coherence length $\xi_0\equiv v_F/\Delta,$ while for energies above the gap $e^{iqR}$ oscillates without decay. These non-vanishing oscillations physically represent quasi-particle propagation in the continuum of the reservoirs, which is therefore not bound by the superconducting coherence length. On the other hand, the phase $k_F R$ oscillates rapidly at the scale of the Fermi wavelength $\lambda_F=2\pi/k_F,$ since the superconducting coherence length is typically much larger than the Fermi wavelength, $\xi_0\simeq 10^3\lambda_F.$ The former length-scale is coupled with the quasi-particle energy $\omega$, while the latter would give a geometric effect. Since the two scales are very different, and we want to focus on new physics related to the energy-dependence rather than to any geometric effects, we will assume that the phase $k_F R$ is fixed. We discuss this choice in some more detail in the Supplementary Material.\\
In what follows, Green's functions always appear in the combination $\sigma_z \tilde{g} \sigma_z,$ $\sigma_z=\smqty(1 & 0 \\ 0 & -1)$ being the Pauli matrix. We will drop the tilde by denoting this combination as $g\equiv\sigma_z \tilde{g} \sigma_z$ for brevity.\\
Introducing the Nambu spinor $$\Psi_m=(u_m(1), v_m(1), u_m(2), v_m(2))^T$$ which collects the amplitudes on the two dots, we can rewrite Eq. (\ref{FLS}) by defining a linear operator $\mathcal{L}$ which acts on the states $\Psi_m$ in the following way:
\begin{equation}\label{tb}
(\mathcal{L}\Psi)_m \equiv M^0_m\Psi_m -M^+_{m+1}\Psi_{m+2}-M^-_{m-1}\Psi_{m-2}=S_m.
\end{equation}
Equation (\ref{tb}) is the Floquet chain advertised above. In the tight-binding analogy, the matrix $M^0_m$ describes a self-energy at position $m$ of the chain, while matrices $M^{\pm}_{m\pm 1}$ describe `hopping' to neighboring sites $m\pm 2.$ The fact that $m$ couples to $m\pm 2$ is a consequence of coupling via second-order Andreev reflection processes. Explicitly, the matrix  $M^0,$
\begin{equation}
    M^0_m =E_m\mathds{1}_4-\mqty(\Sigma_1(m) & g_c(m,R) \\\ g_c(m,R) &\Sigma_2(m)),
\end{equation}
contains a non-local coupling of the two dots through the Green's function of the middle superconductor $g_c(m,R),$ and local Andreev reflection terms on each of the dots, collected in the block-diagonal in the $\Sigma_{1,2}$ matrices:
\begin{equation}
	\begin{split}
	\Sigma_1(m) &=g_c(m)+\mqty(g_a^{11}(m-1) & 0 \\ 0 & g_a^{22}(m+1)),\\
	\Sigma_2(m) &=g_c(m)+\mqty(g_b^{11}(m+1) & 0 \\ 0 & g_b^{22}(m-1)).
	\end{split}
\end{equation}
The matrices $M^{\pm}$ are
\begin{equation}
	\begin{split}
		M^{-}_m &=\mqty*(0 & g^{12}_a(m) & 0 & 0 \\ 0 & 0 & 0 & 0 \\ 0 & 0 & 0 & 0\\ 0 & 0 & g^{21}_b(m) & 0),\\
		M^{+}_m &=\mqty*(0 & 0 & 0 & 0\\ g^{21}_a(m) & 0 & 0 & 0 \\ 0 & 0 & 0 & g^{12}_b(m) \\ 0 & 0 & 0 & 0).
	\end{split}
\end{equation}
Remark: Eq. (\ref{FLS}) is an in-homogeneous equation because of the source terms on the right hand side, collected in the column matrix $S_m.$ In the following sections we are interested in the spectrum on the dots. We can then simply consider the homogeneous version of Eq. (\ref{FLS}), since it is sufficient to find the resonances of the operator $\mathcal{L}.$ In later sections, however, we will be interested in the transport properties (current). Then, we will have to restore the source term. 

\subsection{Iterative Construction of the Resolvent Operator}

\begin{figure*}
	\centering
	\includegraphics[width=0.8\linewidth]{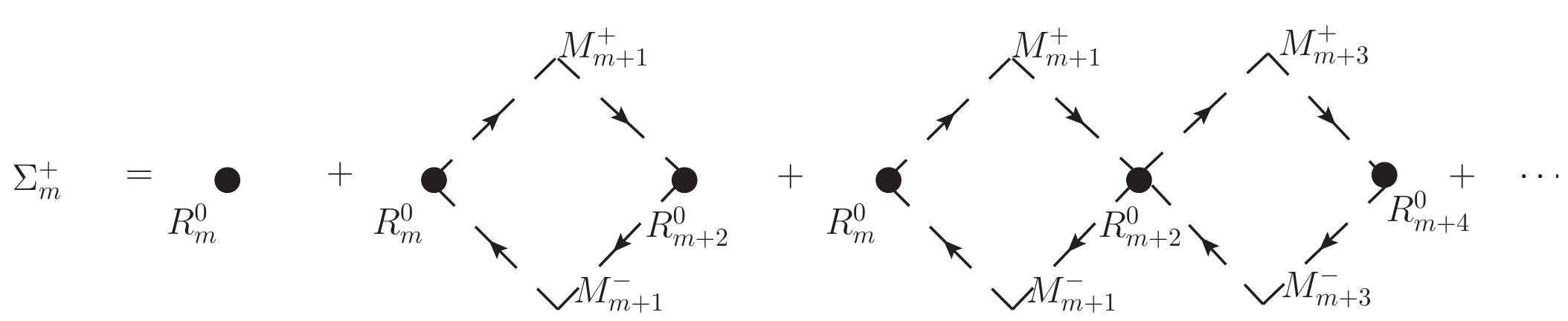}
	\caption{Diagrammatic representation of the forward scattering self-energy. $\Sigma_m^+$ resums loops to the right of site $m.$}
	\label{diagrammatic}
\end{figure*}

If the resolvent operator $\mathcal{R}$ is the inverse of the operator $\mathcal{L}$, then knowledge of  $\mathcal{R}$ allows straightforward calculation of the amplitudes on the dots:
\begin{equation}\label{amplitudes}
	\begin{split}
		u_m(i;lk)=\sum_{i'}&\Bigg[\mathcal{R}_{m,-s_l}^{e_ie_{i'}}J_l(x_{i'}) x_{lk}e^{i\phi_l/2} \\
		&\qquad-\mathcal{R}_{m,s_l}^{e_i h_{i'}}J_l(x_{i'}) y_{lk}e^{-i\phi_l/2}\Bigg],\\
	    v_m(i;lk)=\sum_{i'}&\Bigg[\mathcal{R}_{m,-s_l}^{h_i e_{i'}}J_l(x_{i'}) x_{lk}e^{i\phi_l/2} \\
	    &\qquad-\mathcal{R}_{m,s_l}^{h_i h_{i'}}J_l(x_{i'}) y_{lk}e^{-i\phi_l/2}\Bigg].
	\end{split}
\end{equation}
Finding the poles of the resolvent operator corresponds to finding the spectrum of the operator $\mathcal{L}.$ The resolvent $\mathcal{R}$ is an operator which lives in the extended $\mathrm{dot}\otimes \mathrm{Nambu}\otimes\mathrm{Floquet}$ space. Upper indices correspond to the combined $\mathrm{dot}\otimes \mathrm{Nambu}$ space which is thus of dimension $2\times 2$, and lower indices correspond to the infinite Floquet space. 
The in-homogeneous FLS equations for the resolvent elements are:

\begin{equation}\label{resolvent}
	M^0_m \mathcal{R}_{mn}-M^+_{m+1}\mathcal{R}_{m+2,n}-M^-_{m-1}\mathcal{R}_{m-2,n}=\delta_{mn} \mathds{1} .
\end{equation}

For a tridiagonal block-matrix Hamiltonian, such as the one we are dealing with, it follows generally that its resolvent can be written in continued fraction form \cite{tridiagonal,cf1}. The continued fraction representation is equivalent to the usual Dyson equation \cite{cf2}. 

Starting from Eq. (\ref{resolvent}) it is straightforward to construct the resolvent elements by iteration, assuming a source at some index $n$ and a cutoff at some large Floquet index $\pm N,$ with $\abs{N}\geq \frac{\Delta}{\omega_0}.$ The latter is equivalent to assuming that the wavefunction on the dot decays exponentially at energies above the gap $\abs{\omega+N\omega_0}>>\Delta.$ Physically, the first values of $m$ will correspond to multiple quasi-particle reflection processes, by which the quasi-particle gains energy equal to $m\omega_0$. When $m$ is large enough so that $\omega+m\omega_0>\Delta,$ the quasi-particle enters the superconducting continuum, thus macroscopically resulting into a dissipative quasi-particle flow with normal conductance values. The smaller the voltage value, the more Floquet harmonics we need to take into account. We have the following system consisting of $2N+1$ equations:

\begin{equation}
	\begin{split}
		&M^0_{-N}\mathcal{R}_{-N,n} - M^+_{-N+1}\mathcal{R}_{-N+2,n} - \xcancel{M^-_{-N-1}\mathcal{R}_{-N-2,n}} = 0\\
		& \vdots \\
		&M^0_{n-2}\mathcal{R}_{n-2,n} -M^+_{n-1}\mathcal{R}_{nn}-M^-_{n-3}\mathcal{R}_{n-4,n}=0 \\
		&M^0_n \mathcal{R}_{nn} -M^+_{n+1} \mathcal{R}_{n+2,n} -M^-_{n-1} \mathcal{R}_{n-2,n} = \mathds{1} \\
		&M^0_{n+2} \mathcal{R}_{n+2,n} -M^+_{n+3} \mathcal{R}_{n+4,n}-M^-_{n+1} \mathcal{R}_{nn}=0 \\
		& \vdots \\
		& M^0_N \mathcal{R}_{Nn} -\xcancel{M^+_{N+1} \mathcal{R}_{N+2,n}} -M^-_{N-1}\mathcal{R}_{N-2,n}=0
	\end{split}
\end{equation}

We solve this system of equations by iteration, and find that the diagonal elements are resummed into a geometric series

\begin{equation}\label{Rdiagonal}
	\mathcal{R}_{mm} =\bqty{M^0_m-M^{+}_{m+1}\Sigma^{+}_{m+2}M^{-}_{m+1}-M^{-}_{m-1}\Sigma^{-}_{m-2}M^{+}_{m-1}}^{-1}
\end{equation}

with forward and backward self-energy matrices, $\Sigma^{\pm},$ that can be calculated recursively once boundary conditions are imposed, that is once $\Sigma^{\pm}_{\pm N}$ is set to 0 at some large number $\pm N.$ We find that

\begin{equation}\label{self-energies}
	\begin{split}
		&\Sigma^{+}_m=\frac{1}{M^0_{m}-M^{+}_{m+1}\Sigma^{+}_{m+2}M^{-}_{m+1}}, \\
		&\Sigma^{-}_m=\frac{1}{M^0_{m}-M^{-}_{m-1}\Sigma^{-}_{m-2}M^{+}_{m-1}}.
	\end{split}
\end{equation}
\\
The non-diagonal resolvent elements $\mathcal{R}_{mn}$ can then be expressed using the self-energy matrices

\begin{equation}\label{paths}
	\begin{split}
		&\mathcal{R}_{mn} =\Sigma^{+}_m M^{-}_{m-1}\dots \Sigma^{+}_{n+2} M^{-}_{n+1}\mathcal{R}_{nn} \qif{m>n}, \\
		&\mathcal{R}_{mn} =\Sigma^{-}_m M^{+}_{m+1}\dots \Sigma^{-}_{n-2} M^{+}_{n-1}\mathcal{R}_{nn} \qif{m<n}.
	\end{split}
\end{equation}

This description of the resolvent admits a simple diagrammatic representation. For example, by expanding the forward self-energy term $\Sigma^+$ into a series we see that this term regroups all paths that start from a point $m$ and only return to it after having visited all sites $m'>m,$ up to the boundary site $N.$

\begin{equation}
	\begin{split}
		\Sigma^{+}_m &=\frac{\mathcal{R}^0_m}{1-\mathcal{R}^0_m M^{+}_{m+1}\Sigma^{+}_{m+2}M^{-}_{m+1}} \\
		&= \mathcal{R}^0_m+ \mathcal{R}^0_m M^{+}_{m+1}\Sigma^{+}_{m+2}M^{-}_{m+1}\mathcal{R}^0_m+\dots \\
		&= \mathcal{R}^0_m+ \mathcal{R}^0_m M^{+}_{m+1}\mathcal{R}^0_{m+2}M^{-}_{m+1}R^0_m+\dots	
	\end{split}
\end{equation}

The expansion of the forward self-energy in terms of diagrams is illustrated in Fig. \ref{diagrammatic}. 
On the other hand, the backward self-energy term $\Sigma^-$ resums loops which pass through sites $m'<m$. Equation (\ref{paths}) then describes the shortest path connecting a site $n$ to site $m.$ The role of the self-energy terms $\Sigma^{\pm}$ is to renormalize the unperturbed diagonal elements of the resolvent $\mathcal{R}^0_m=1/M^0_m$ by introducing a finite imaginary part, corresponding to virtual excursions to the superconducting reservoirs. \par
This expansion is a locator-type expansion of the resolvent, commonly used in disordered systems \cite{anderson,ziman}, in the sense that the unperturbed part of the resolvent $\mathcal{R}^0_m$ locates a quasi-particle on site $m,$ in contrast to the more usual `propagator' which describes the propagation of a free particle. The resolvent $\mathcal{R}_{mn}$ then represents the probability that a quasi-particle is localized on site $m,$ given that it was originally on site $n.$ The major difference between a propagator and a locator expansion is the restriction on repeated indices which is necessary in the later.

\subsection{Current}
Given that the dressed quasi-particle operators $\Gamma^{\dagger},\Gamma$ form a complete basis, we can express all other operators in this basis. The advantage of such decompositions is that one can then very easily derive expressions for expectation values in the stationary state $\ket{S}$, which is simply defined as the state which is annihilated by the application of the $\Gamma$ operator: $\Gamma_{lk\sigma}\ket{S}=0.$

In general, the current from a dot $i$ to a reservoir $j$ is given by 
\begin{equation}
    	I_{i \to j}(t)=-i\pqty{e^{-is_j\omega_0 t}\psi_{ji\sigma}^{\dagger}(t)d_{i\sigma}(t) - e^{is_j\omega_0 t}d_{i\sigma}^{\dagger}(t)\psi_{ji\sigma}(t)}
\end{equation}
where the shorthand $\psi_{ji\sigma}^{\dagger}= \sum_k J_j^{\star}(x_i)c_{jk\sigma}^{\dagger}$ is used.
Creation and annihilation operators on the dots and on the leads can be expressed as functions of the dressed Floquet operators $\Gamma$
\begin{widetext}
\begin{equation}
	\begin{split}
		d_{i\sigma}(t) &=\sum_{l,k,m} \pqty{e^{-i(E_{lk}+m\omega_0)t} u_m(i;lk) \Gamma_{lk\sigma} -\sigma e^{i(E_{lk}+m\omega_0)t}v^{\star}_m(i;lk)\Gamma_{lk-\sigma}^{\dagger}},\\
		\psi_{ji\sigma}^{\dagger}(t) &= \sum_{l,k,m}\Big[ J_j^{\star}(x_i) e^{i(E_{lk}+m\omega_0)t} \pqty{\delta_{m0} \delta_{jl} x_{jk}e^{-i\phi_j/2}+U^{\star}_m(j;lk)} \Gamma_{lk\sigma}^{\dagger} \\
		& \qquad \qquad-\sigma J_j(x_i)e^{-i(E_{lk}+m\omega_0)t} \pqty{\delta_{m0} \delta_{jl}  y_{jk}e^{-i\phi_j/2}+V_m(j;lk)} \Gamma_{lk-\sigma}\Big].
	\end{split}
\end{equation}

Since the steady state is the state which is annihilated by the operator $\Gamma,$ any average taken in the steady state will contain only contributions from $\expval{\Gamma \Gamma^{\dagger}}{S}$ terms. We then find that the steady-state current from dot $i$ to reservoir $j$ is	
\begin{equation}
	\expval{I_{i\to j}(t)}= 4\Im\sum_{l,k} \sum_{m,n} \bqty{J_j(x_i) e^{-i(m-n)\omega_0t} \pqty{\delta_{m,s_j} \delta_{jl} y_{jk}e^{-i\phi_j/2}+V_{m-s_j}(j;lk)}v^{\star}_n(i;lk)}.
\end{equation}
\end{widetext}
Using the FLS equations (\ref{FLS}), we can re-express the terms involved on the RHS so that the current contains only quadratic terms in the resolvent. First, we re-express the source term:
\begin{equation}
\begin{split}
J_j(x_i)\delta_{m,s_j}y_{jk}&e^{-i\phi_j/2} =-(E_{jk}+m\omega_0)v_m(i;jk) \\
&+\sum_{l,i'}\Bigg[g_{l,ii'}^{21}(m-s_l) u_{m-2s_l}(i';jk) \\ 
&\qquad\quad+ g_{l,ii'}^{22}(m-s_l) v_m(i';jk)\Bigg],
\end{split}
\end{equation}
and the amplitudes in the reservoirs
\begin{equation}
\begin{split}
V_j(m-s_j)= \sum_{i'}&\Bigg[g_{j,ii'}^{21}(m-s_j) u_{m-2s_j}(i';l) \\
&\qquad- g_{j,ii'}^{22}(m-s_j) v_m(i';l)\Bigg].
\end{split}
\end{equation}
The expressions we find include terms which are local $g_{l,ii},$ and non-local $g_{l,ii'}$ in the Green's functions, as well as in the resolvent (the non-local terms in the resolvent are its non-diagonal blocks). Note that only $g_{c,i\neq i'}\neq 0$ since only the middle superconductor $S_c$ connects the two dots. We can therefore categorize contributions in these expressions by the number of non-local quantities they contain. A first approximation is to keep the ''0-th" order terms, containing only local contributions. Then, the current from the dot labeled $1$ to the middle superconductor is given by
\begin{equation}\label{twodoteq}
	\begin{split}
		\expval{I_{\mathrm{1}\to c}}_{\mathrm{dc}}^0 &=4\Im\int_{\Delta}^{+\infty} \dd\omega \sum_{m} \bigg[\mqty(g^{21}_c(m), & -g^{22}_c(m)) \\ 
		&\qquad\times \mqty(\mathcal{R}^{11}_{m,1} & \mathcal{R}^{12}_{m,-1} \\ \mathcal{R}^{21}_{m,1} & \mathcal{R}^{22}_{m,-1})Q_a \mqty(\mathcal{R}^{21}_{m,1} \\ \mathcal{R}^{22}_{m,-1})^{\star} \\
		&-\mqty(g^{21}_a(m+1), & -g^{22}_a(m+1)) \\ 
		&\qquad\times \mqty(\mathcal{R}^{11}_{m+2,0} & \mathcal{R}^{12}_{m+2,0} \\ \mathcal{R}^{21}_{m,0} & \mathcal{R}^{22}_{m,0})Q_c \mqty(\mathcal{R}^{21}_{m,0} \\ \mathcal{R}^{22}_{m,0})^{\star} \bigg].
	\end{split}
\end{equation}
Note that this is exactly the result one would find in the case of a single junction. The resolvent elements however, are calculated by resuming the contribution of different paths, as explained in the preceding section, and therefore take into account the effects due to the proximity of a second dot. The matrices $Q_l(\omega)$ describe the populations in the reservoirs of ejection. For the current we will take integrals over excitation energies above the gap so that $\omega>\Delta.$ Then, we find that
\begin{equation}
	Q_l(\omega)=\frac{2\Gamma_l}{\sqrt{\omega^2-\Delta^2}}\mqty(\omega & -\Delta \\ -\Delta^{\ast} & \omega),\qquad \omega>\Delta.
\end{equation}
If we consider that the quasi-particle density of states is \cite{coleman} $\rho_S(\omega)=2\rho_0\frac{\abs{\omega}}{\sqrt{\omega^2-\Delta^2}}\theta(\abs{\omega}-\Delta),$ where $\rho_0$ is the density of states in the normal state, we see directly that the diagonal of $Q$ is nothing other than $\pi J^2\rho_S(\omega).$ This translates the fact that the MAR current will naturally depend on the populations of the reservoirs.

Equation (\ref{twodoteq}) is at first sight not easy to calculate since it involves an integral over all quasi-particle excitation energies above the gap up to infinity, as well a summation over the Floquet harmonics. Fortunately, the resolvent elements decay exponentially at large energies, so that the integration can be drastically truncated. Moreover, at large enough voltages we observe a localization (analogous to the Wannier-Stark localization) which gives a rapidly convergent summation over the Floquet harmonics. These points will be further discussed in Sec. IV. Additional figures showing the localization of the resolvent elements are presented in the Supplementary Material.

\section{Spectroscopy: Revealing the Floquet-Andreev ladders}\label{spectroscopy}
\begin{figure}
	\centering
	\begin{subfigure}{0.46\linewidth}
		\centering
		\includegraphics[width=\textwidth]{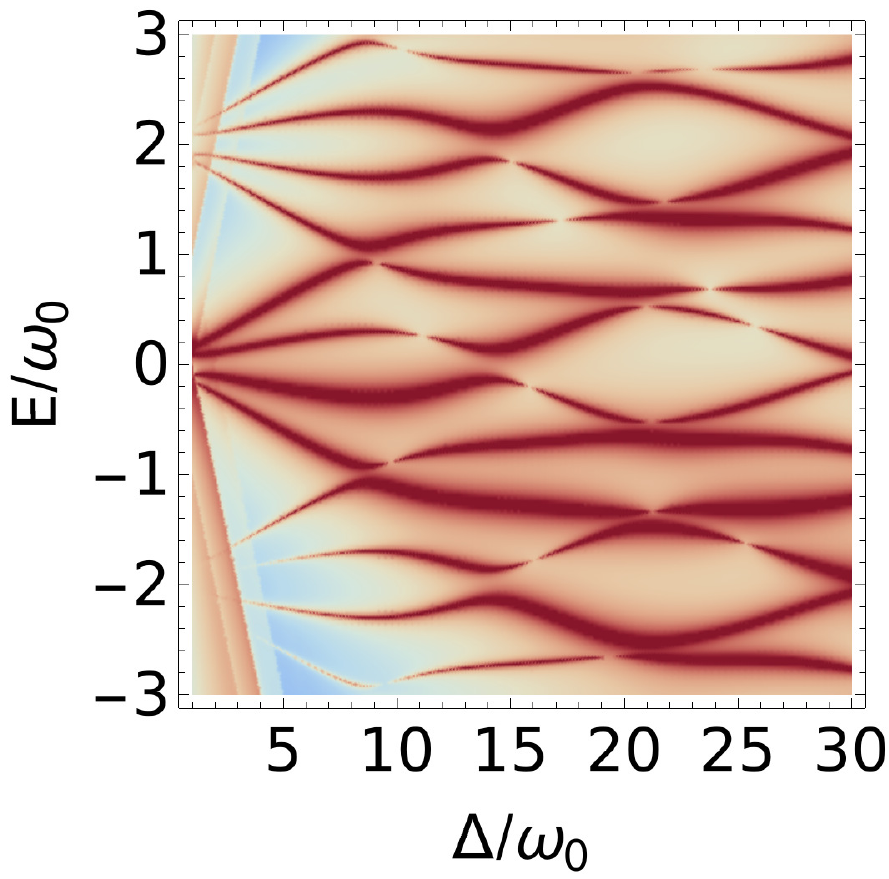}
		\caption{$\Gamma_{a,b,c}=0.1\Delta, R=\xi_0/2$}
	\end{subfigure}
	\hfill
	\begin{subfigure}{0.46\linewidth}
		\centering
		\includegraphics[width=\textwidth]{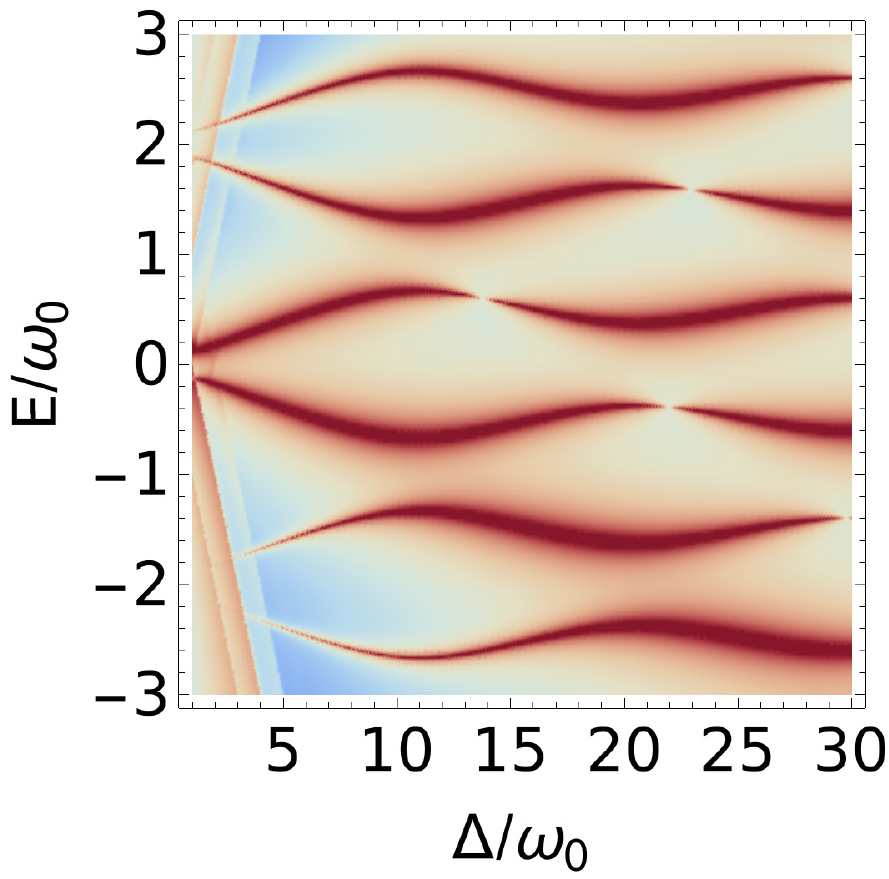}
		\caption{$\Gamma_{a,b,c}=0.1\Delta, R=5\xi_0$}
	\end{subfigure}
	\begin{subfigure}{0.46\linewidth}
		\centering
		\includegraphics[width=\textwidth]{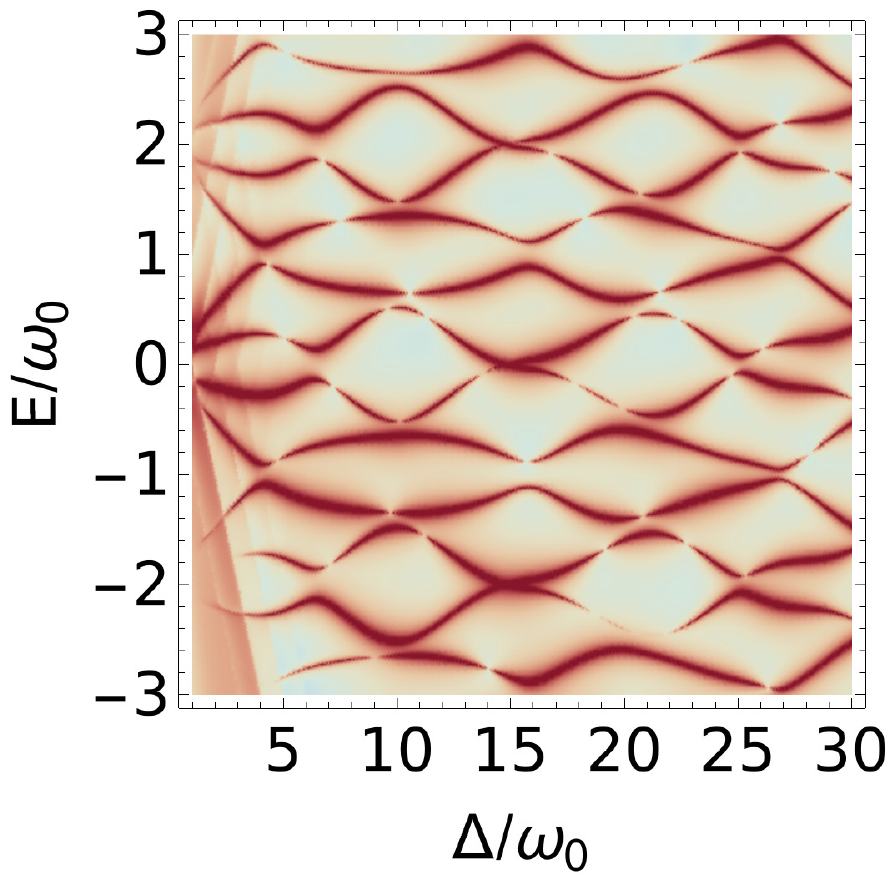}
		\caption{$\Gamma_{a,b,c}=0.3\Delta, R=\xi_0/2$}
	\end{subfigure}
	\hfill
	\begin{subfigure}{0.46\linewidth}
		\centering
		\includegraphics[width=\textwidth]{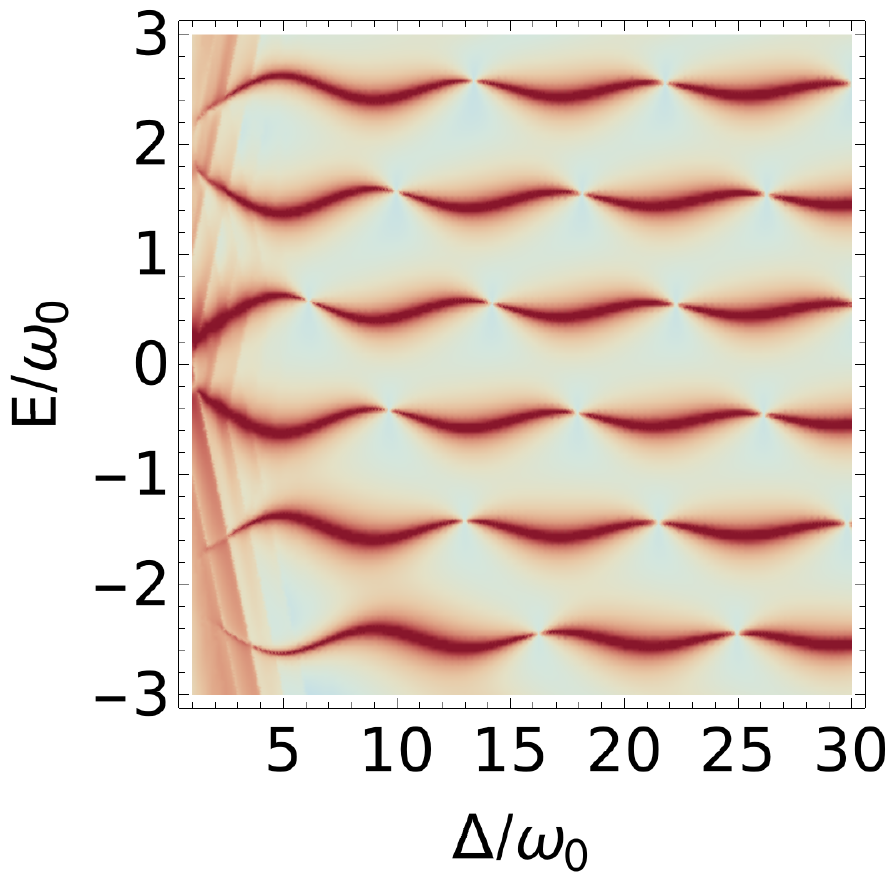}
		\caption{$\Gamma_{a,b,c}=0.3\Delta, R=5\xi_0$}
	\end{subfigure}
	\begin{subfigure}{0.46\linewidth}
		\centering
		\includegraphics[width=\textwidth]{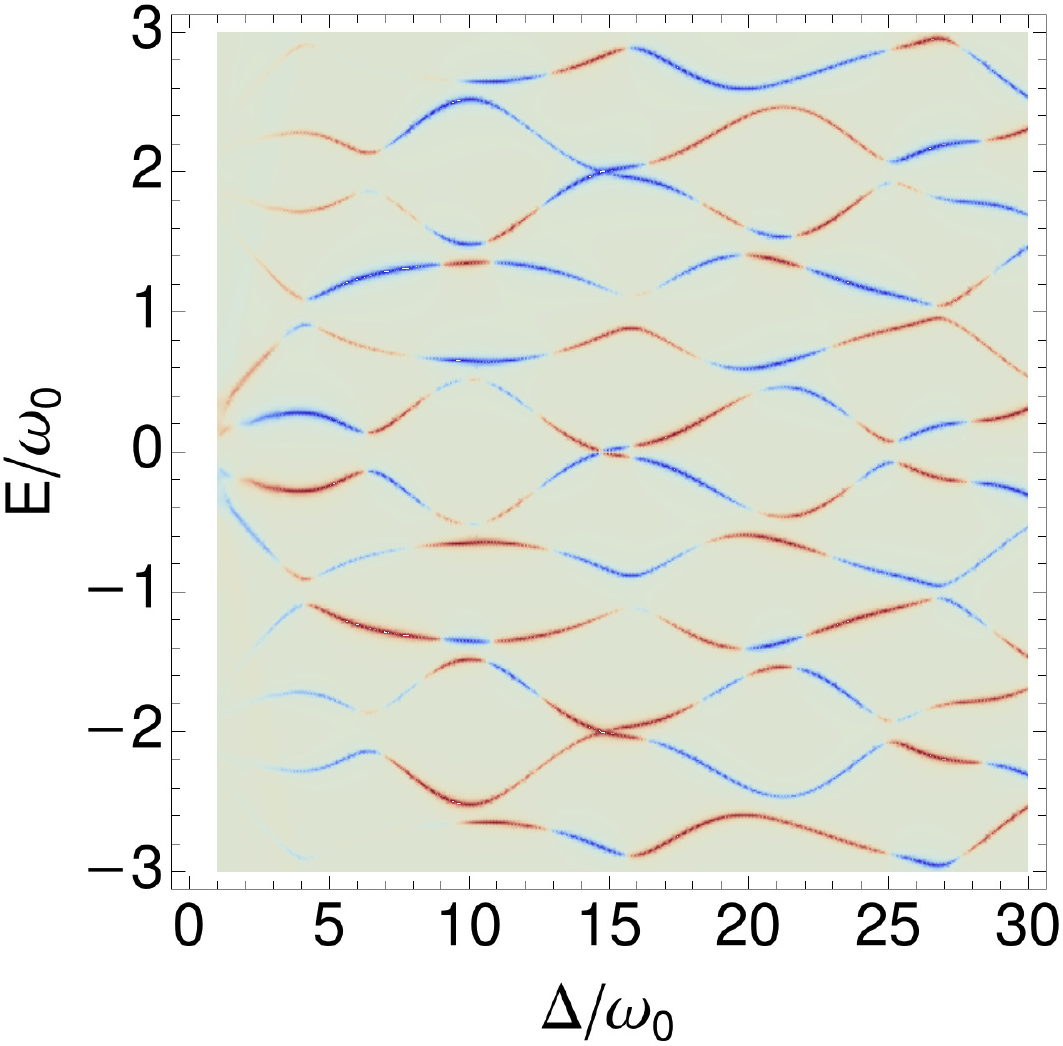}
		\caption{$\Gamma_{a,b,c}=0.3\Delta, R=\xi_0/2$}
	\end{subfigure}
	\hfill
	\begin{subfigure}{0.46\linewidth}
		\centering
		\includegraphics[width=\textwidth]{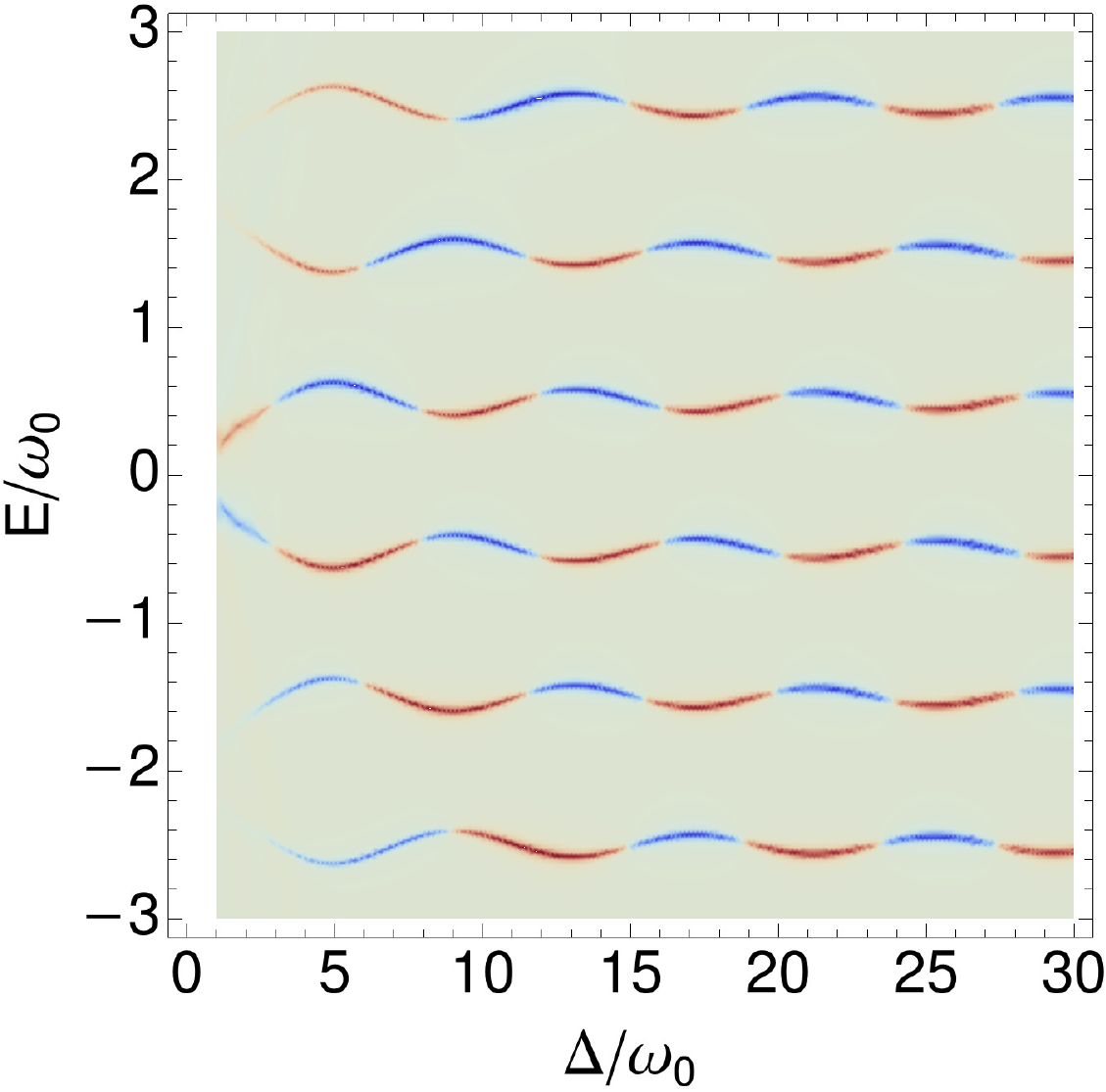}
		\caption{$\Gamma_{a,b,c}=0.3\Delta, R=5\xi_0$}
	\end{subfigure}
	\caption{(a-d) Density plots of the spectral function on the first dot, $-\frac{1}{\pi} \Im \Tr_{\mathrm{dot 1}} \mathcal{R}_{00},$ plotted using inverse scaling. Red color corresponds to the maxima of the spectral function. The distance between the dots is taken to be $R=\xi_0/2$ on (a,c,e)  and $R=5\xi_0$ on (b,d,f). (e-f) Particle-hole asymmetry of the spectral function $-\frac{1}{\pi}\Im[\mathcal{R}^{11}_{00}(\omega)-\mathcal{R}^{22}_{00}(\omega)].$ Red represents positive values of the difference (electron-like) and blue represents negative values (hole-like). The signs change at avoided crossings.}
	\label{bijunction spectrum}
\end{figure}
The diagonal part of the resolvent in Floquet space gives access to a spectral function. Indeed, in the case of Floquet-Green functions, one can still define a spectral function which can be interpreted as a density of states \cite{positivity}. The quantity
\begin{equation}
	\mathcal{A}(\omega)=-\frac{1}{\pi}\Im \mathcal{R}_{00}(\omega)
\end{equation}
can be seen as a time-average of the spectral function over one period of the drive. Whether we need to take the trace over the Nambu subspace of one dot or not should depend on the type of spectroscopy experiment one performs. For example, if we perform a local tunneling spectroscopy measurement on one dot, we can probe both the creation and destruction of excitations, while in an ARPES experiment we can only extract electrons, so we will only have access to one part of the spectrum \cite{thesispillet}. In the case of local tunneling spectroscopy with a normal probe coupled to the first dot, for example, the spectral function will be given by a trace over the subspace of the first dot, defined as follows:
\begin{equation}\label{spectral function}
\begin{split}
\mathcal{A}_{\mathrm{dot 1}}(\omega) &=-\frac{1}{\pi}\Im\Tr_{\mathrm{dot 1}} \mathcal{R}_{00}(\omega) \\ &=-\frac{1}{\pi}\Im[\mathcal{R}^{11}_{00}(\omega)+\mathcal{R}^{22}_{00}(-\omega)].
\end{split}
\end{equation}
\begin{figure*}
	\centering
	\includegraphics[width=\linewidth]{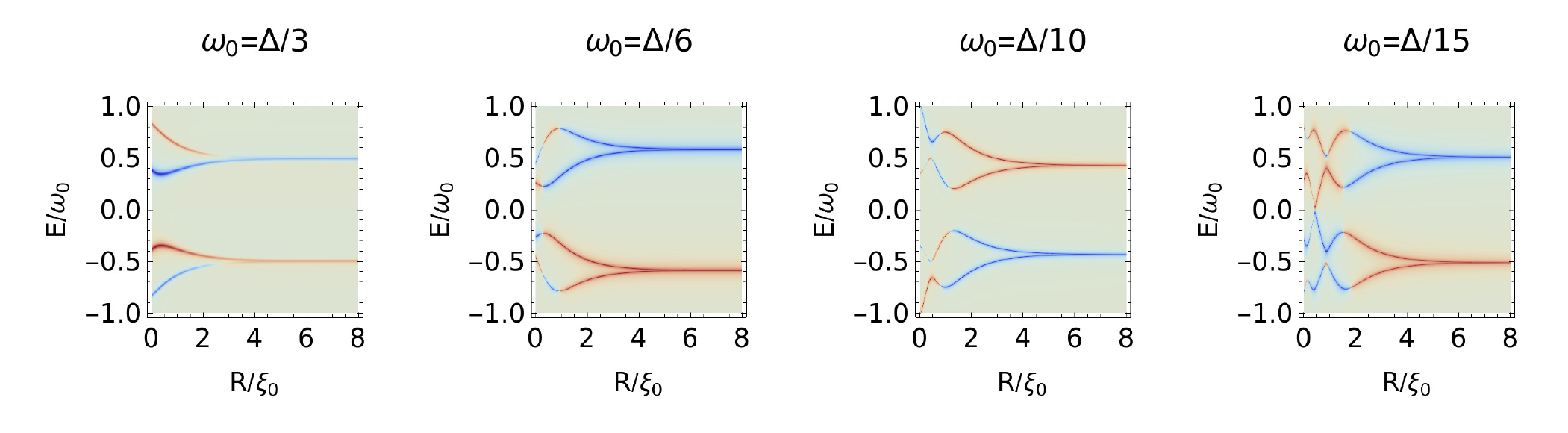}
	\caption{Density plots of $-\frac{1}{\pi}\Im[\mathcal{R}^{11}_{00}(\omega)-\mathcal{R}^{22}_{00}(\omega)]$ as a function of the distance between the two dots, for different values of voltage bias $\omega_0$. At large distances $R\gg \xi_0$ the quasienergies approach those of a single junction. At $R\sim \xi_0$ we are in a molecular regime, and there is a considerable lift of degeneracy of the quasinergies. The intensity of the red (positive values) and blue lines (negative values) should not be compared across plots, as it is varied in each plot to achieve the best visibility. We set $k_FR=\pi/4$ and $\Gamma_{a,b,c}=0.3\Delta.$}
	\label{splitting}
\end{figure*}
The resolvent is calculated using the iterative formula of Eq. (\ref{Rdiagonal}). In practice, we found that a cutoff index of the order of $N\sim\frac{\Delta}{\omega_0}$ is sufficient for convergence when calculating the spectral plots. Since the system we are studying is periodic in time with a period $T=\frac{2\pi}{\omega_0},$ we can plot quantities as functions of inverse voltage $\frac{1}{\omega_0}$ in order to make apparent the periodicity.
\paragraph{Floquet spectrum.} Figure \ref{bijunction spectrum} presents the spectral function as defined in Eq. (\ref{spectral function}), calculated at a fixed distance between the two dots. At large distances compared to the superconducting length $R>\xi_0,$  the energies on the first dot tend to that of a single junction, and we obtain results which agree with previous calculations done using the method of Keldysh Green's functions \cite{engineering,fwsa}. The spectrum of the single resonant dot consists of ladders of resonances at quasi-energies $E_m=\epsilon_{\pm}+2m\omega_0,$ with $\epsilon_{+}=-\epsilon_{-}$ \cite{engineering} and has a basic period of $2eV.$ The Floquet ladders show avoided crossings which is a sign of coupling between them via Landau-Zener transitions. Both electronic and hole parts of the resolvent have peaks at the same energies, but their weight away from avoided crossings differs when varying the voltage. To illustrate this, we plot the difference between the electron and hole parts of the spectral function $-\frac{1}{\pi}\Im[\mathcal{R}^{11}_{00}(\omega)-\mathcal{R}^{22}_{00}(\omega)]$ in Fig. \ref{bijunction spectrum}(e-f). Near avoided crossings, a rapid change of the Floquet states is expected to happen. Accordingly, we see that the sign of the aforementioned quantity changes signs at avoided crossings, signaling the change in character between electron-like and hole-like states. In order to observe this asymmetry it would be necessary to break the mirror symmetry of the system, for example by having asymmetric tunnel couplings. One could then probe the asymmetry away from avoided crossings by measuring the conductance at opposite voltage values, as has been proposed in reference \cite{asymmetry}. 
\paragraph{Floquet engineering.} The avoided crossings present in Fig. \ref{bijunction spectrum} could be used to find dynamical sweet spots of the system. These so-called sweet spots are optimal working points corresponding to the extrema in quasi-energy differences and have been proposed as a way to protect qubits from noise. Contrary to the static case where few sweet spots are present, the extra dimension of time in periodically driven systems allows to find a manifold of dynamical sweet spots \cite{sweetspots}. An added advantage is that one can tune the system to an avoided crossing by changing the drive in situ. Realization of a Floquet qubit has been proposed along this line \cite{floquetqubit}, where a periodic driving can be used to tune the system near one of the avoided crossings, and a second drive can be used to control transitions between the Floquet states. Typically, the smaller the quasi-energy dispersion, the more insensitive would the qubit states be to fluctuations. Moreover, tuning a fluxonium qubit to a dynamical sweet spot away from its  half-flux bias static spot has been shown to increase coherence times \cite{fluxonium}, demonstrating the relevance of Floquet engineering to the qubit community. Some more Floquet spectra for different combinations of tunnel couplings are presented in the Supplementary Material, showing various different possibilities when engineering the band-structure.
\paragraph{Level splitting.} When the distance between the dots is comparable to the superconducting coherence length $R\sim\xi_0,$ we arrive at an `Andreev molecule regime',  where a lift of degeneracy produces four peaks instead of two in each Floquet-Brillouin zone. At large distances the splitting decreases exponentially with the distance $\sim e^{-R/\xi_0}.$ However, at intermediate distances its behavior depends on the applied voltage. Figure \ref{splitting} shows the lift of degeneracy of the quasienergy levels when the two dots are brought close together. In reality, we observe \textit{oscillations} of the spectral functions around the single-junction value even at very large distances. We discuss this long-range interference effect in Sec. IV.B. and show that the oscillations of the spectral functions lead to an oscillatory I--V curve. 

\section{Subgap Current}\label{current}
Before showing the results for the Andreev molecule, we will briefly discuss the well-known case of a single junction. This serves two purposes, the first being to benchmark our method by comparing with what is already known, and, secondly, to acquaint the reader with the Floquet ladder and its impact on the current.

\subsection{Current through one resonant dot from the point of view of its Floquet spectrum}
\begin{figure}
	\centering
	\begin{subfigure}{0.56\linewidth}
		\centering
		\includegraphics[width=\textwidth]{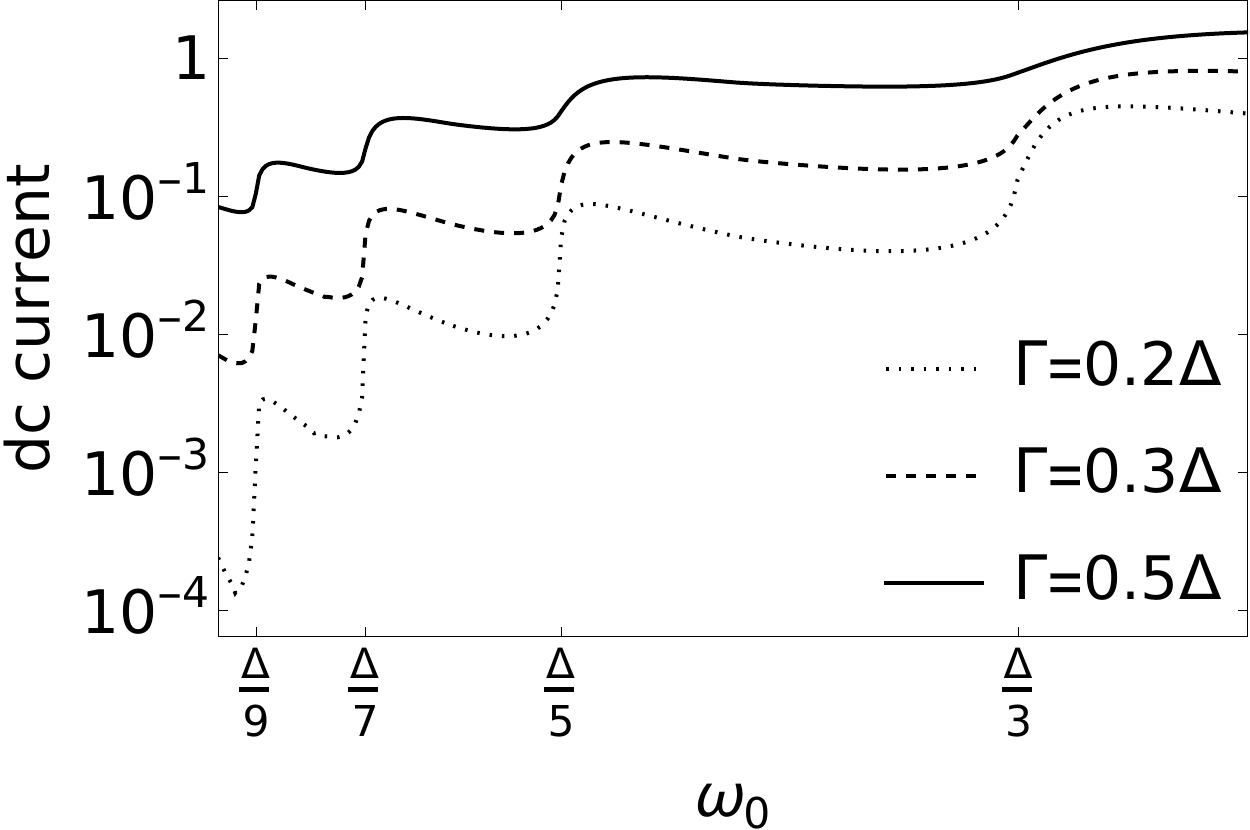}
		\caption{}
	\end{subfigure}
	\hfill
	\begin{subfigure}{0.4\linewidth}
		\centering
		\includegraphics[width=\textwidth]{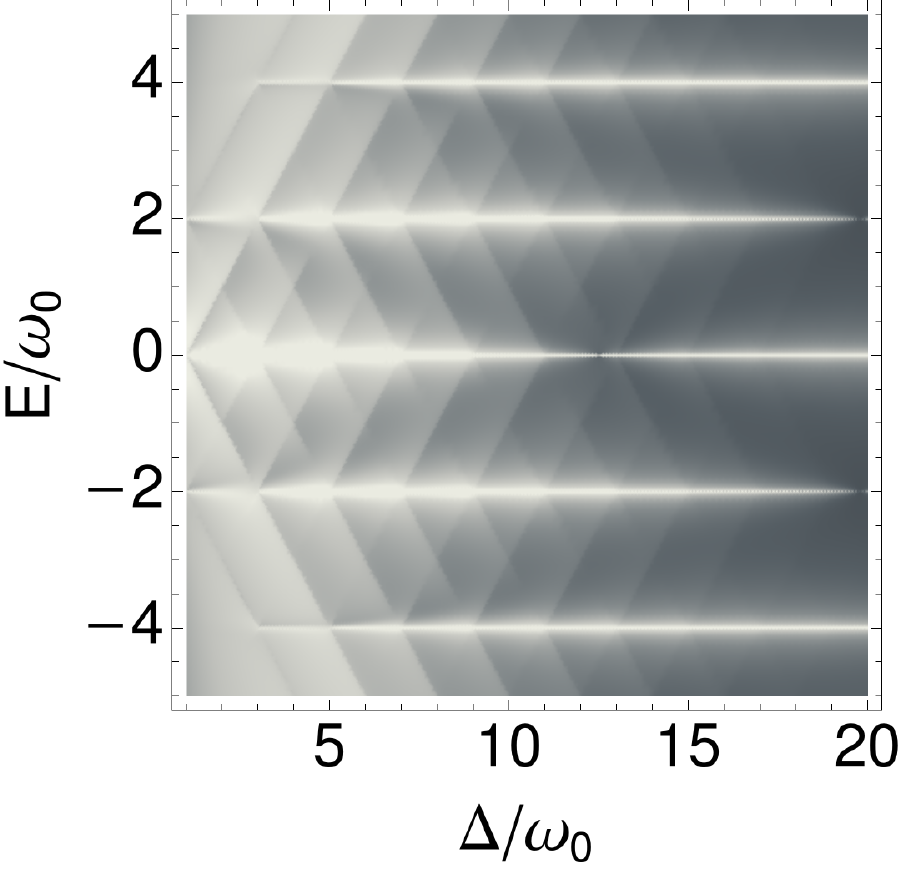}
		\caption{}
	\end{subfigure}
	\begin{subfigure}{0.56\linewidth}
		\centering
		\includegraphics[width=\textwidth]{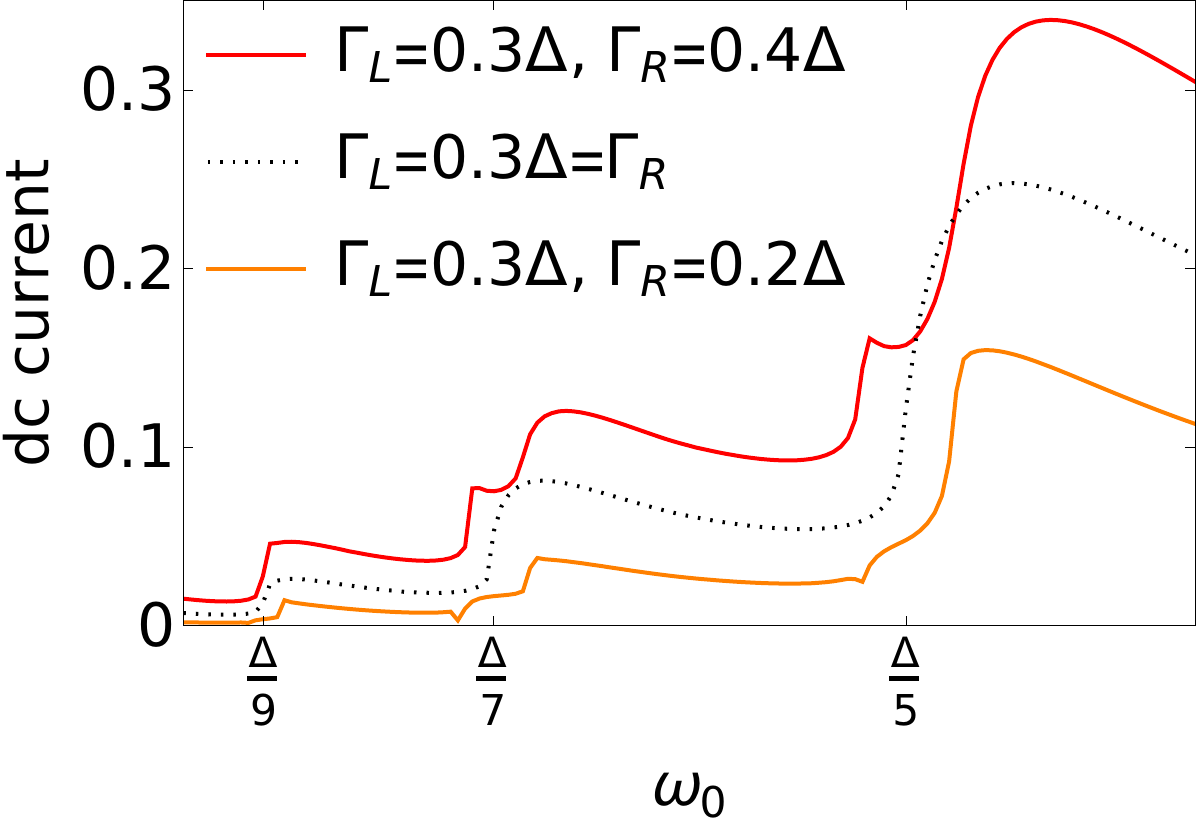}
		\caption{}
	\end{subfigure}
	\hfill
	\begin{subfigure}{0.4\linewidth}
		\centering
		\includegraphics[width=\textwidth]{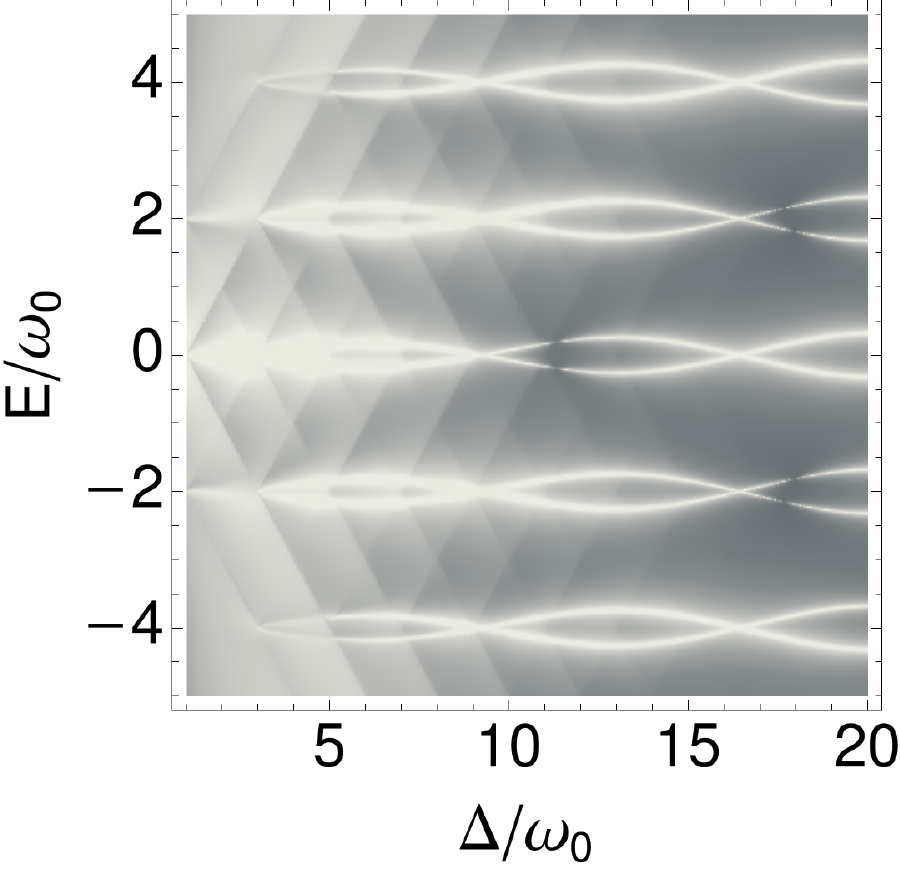}
		\caption{}
	\end{subfigure}	
	\caption{Subgap current and corresponding Floquet spectra of a single resonant dot. (a) Subgap structure for the highly symmetric left-right configuration $\Gamma=\Gamma_L=\Gamma_R$ and $(V_L=-V, V_R=+V)$ with the known MAR steps at odd subdivisions of the gap. Logarithmic scaling has been used for better visibility of the features. (b) The corresponding Floquet spectrum consists of decoupled ladders at multiples of the basic frequency $2\omega_0.$ The maxima of the spectral function are shown here in white. (c) Comparison of subgap current for equal (dashed black line) and unequal (red and orange lines) tunnel couplings ($\Gamma_L\neq\Gamma_R$), showing the modification of the MAR steps when there is asymmetry in the tunnel couplings. (d) Density plot of $-\Im[\mathcal{R}_{00}^{11}(\omega)+\mathcal{R}_{00}^{22}(\omega)]$ for $\Gamma_L=0.3\Delta, \Gamma_R=0.4\Delta.$}
	\label{onedotcurrent}
\end{figure}

The symmetric configuration of a resonant dot with energy $\epsilon_d=0$ coupled to two reservoirs which are voltage biased with $(V_L=-V,V_R=+V)$ is already well understood: the subgap structure of the current-voltage curve shows steps at voltages $2eV=\frac{2\Delta}{n},$ and the presence of the resonant level restricts $n$ to being an odd integer, while even MAR processes are suppressed \cite{QDreview,yeyati-dot,Jonckheere,shumeiko}. These steps in the subgap current appear whenever a new `MAR trajectory' becomes possible: a quasi-particle is Andreev reflected $n$ times, changing its energy by $e(V_R-V_L)=2eV$ with each reflection, until it has enough energy (equivalent to the size of the gap $2\Delta$) to reach the superconducting continuum of states and give a contribution to the current. The results of our calculation for this case are depicted on Fig. \ref{onedotcurrent}(a) for different values of tunnel couplings $\Gamma=\Gamma_L=\Gamma_R$, and we verify that our method gives the expected results for the I--V curves at zero temperature. In order to produce this result, we numerically calculated the current given by the following formula:
\begin{equation}\label{onedot}
	\begin{split}
		\expval{I_{\mathrm{dot}\to R}}_{\mathrm{dc}} &= 4\Im\int_{\Delta}^{+\infty} \dd\omega \sum_{m} \bigg[\mqty(g^{21}_L(m+1), & g^{22}_L(m+1)) \\
		&\qquad\times \mqty(\mathcal{R}^{11}_{m+2,-1} & \mathcal{R}^{12}_{m+2,1} \\ \mathcal{R}^{21}_{m,-1} & \mathcal{R}^{22}_{m,1})Q_R \mqty(\mathcal{R}^{21}_{m,-1} \\ \mathcal{R}^{22}_{m,1})^{\star} \\
		&-\mqty(g^{21}_R(m-1), & g^{22}_R(m-1)) \\
		&\qquad\times \mqty(\mathcal{R}^{11}_{m-2,1} & \mathcal{R}^{12}_{m-2,-1} \\ \mathcal{R}^{21}_{m,1} & \mathcal{R}^{22}_{m,-1}) Q_L \mqty(\mathcal{R}^{21}_{m,1} \\ \mathcal{R}^{22}_{m,-1})^{\star}\bigg].
	\end{split}
\end{equation}
We see that the current requires calculation of the resolvent elements $\mathcal{R}_{m,n}$ which connect the `source sites' at $n=V_{L,R}/V=\pm 1$ on the Floquet chain to sites on positions $m.$ These non-diagonal elements correspond to processes where the system changes its energy by $\abs{m-n}\omega_0,$ equivalent to absorbing or emitting an $\abs{m-n}$ number of `photons'.
\paragraph{Resolvent resonant structure.} In the previous section we have seen that the resolvent $\mathcal{R}_{00}$ has resonances at even multiples of $\omega_0:$ outside the gap the spectrum shows signs of dissipation because of strong hybridization with the reservoirs, i.e. the resolvent decays exponentially for large energies $\omega\gg\Delta.$ Floquet resonances, however, `survive' around frequencies $\omega_0>\Delta/n,$ with $n$ the smallest even number such that the condition holds. The peaks are sharper at small voltages, and become smeared when increasing $\omega_0.$ The symmetry by translation of the resolvent $\mathcal{R}_{m,n}(\omega+p\omega_0)=\mathcal{R}_{m+p,n+p}(\omega)$ means that elements $\mathcal{R}_{\pm 1,\pm 1}(\omega)=\mathcal{R}_{00}(\omega\pm\omega_0)$ have resonance peaks around odd multiples of $\omega_0,$ which gives the condition for the MAR steps at odd subdivisions of the gap in the presence of a resonant level. Moreover, since the non-diagonal elements of the resolvent $\mathcal{R}_{m,\pm 1}$ are obtained from the diagonal elements $\mathcal{R}_{\pm 1,\pm 1}$ through Eq. (\ref{paths}) we can conclude two things: a) they are resonant when the corresponding diagonal elements are resonant, and b) there is a hierarchy of peaks in $m$ which depends on the voltage. An element $\mathcal{R}_{m,\pm 1}$ becomes dominant when $m\omega_0$ is the dominant peak above the gap, i.e. when $m$ is the minimal odd integer for which $m\omega_0>\Delta.$ For example, the element $R_{-3,-1}$ is dominant when the second order MAR trajectory is dominant, $\frac{\Delta}{3}<\omega_0<\Delta,$ the element $R_{-5,-1}$ is dominant when the third order MAR trajectory is dominant, $\frac{\Delta}{5}<\omega_0<\frac{\Delta}{3},$ and so on. The resonant structure of the resolvent is illustrated with some more detail in the Supplementary Material.

The localization on the Floquet chain can be further illustrated by decomposing the subgap current into the contribution from each Floquet harmonic. Then, we find that at large voltages only a few harmonics need to be taken into account when calculating the sum in Eq. (\ref{onedot}), and the number of harmonics needed increases as the voltage goes to zero. We will show, and further discuss, this localization in the case of the Andreev molecule (Fig. \ref{harmonics}), but a completely analogous result holds for the single junction case.
\paragraph{Asymmetry effects.} It has perhaps been less commented in the literature that the subgap current steps appear at exactly $\omega_0=\frac{\Delta}{n}$ only when there is a `left-right' parity symmetry which happens when the tunnel couplings to the reservoirs are equal, $\Gamma_L=\Gamma_R$ and the voltage-biasing is symmetric around the dot energy $V_L=-V_R.$ In this mirror-symmetric case, electron-like and hole-like MAR trajectories are equally favorable. The corresponding spectrum then consists of degenerate ladders situated exactly at even multiples of $\omega_0.$ The Floquet ladders are completely decoupled and show no avoided-crossing behavior, as shown on Fig. \ref{onedotcurrent}(b). When this symmetry is broken, we find that the electron-like part $-\Im\mathcal{R}_{00}^{11}(\omega)$ and the hole-like part $-\Im\mathcal{R}_{00}^{22}(\omega)$ of the spectrum have peaks at different energies $E_m=m\omega_0\pm \epsilon,$ as has been already noted \cite{fwsa,engineering}. We find that the current carries a trace of this characteristic of the spectrum: the MAR steps break into two sub-steps, positioned around the original $\omega_0=\frac{\Delta\pm \epsilon}{n}$ frequencies. The exact shape of the steps (cusps or peaks) depends on the choice of the couplings \cite{PhysRevB.82.060517}, as shown in Fig. \ref{onedotcurrent}(c). This result suggests that electron-boson interaction (such as absorption or emission of photons by the tunneling quasi-particles), as studied in reference \cite{asymmetry2} is not sufficient to break particle-hole symmetry of the conductance, but also requires a breaking of the mirror symmetry of the system. In reference \cite{asymmetry2} the mirror symmetry was broken by considering a N-S system.

\subsection{Current through the driven Andreev molecule}
\begin{figure}
	\centering
	\includegraphics[width=0.9\linewidth]{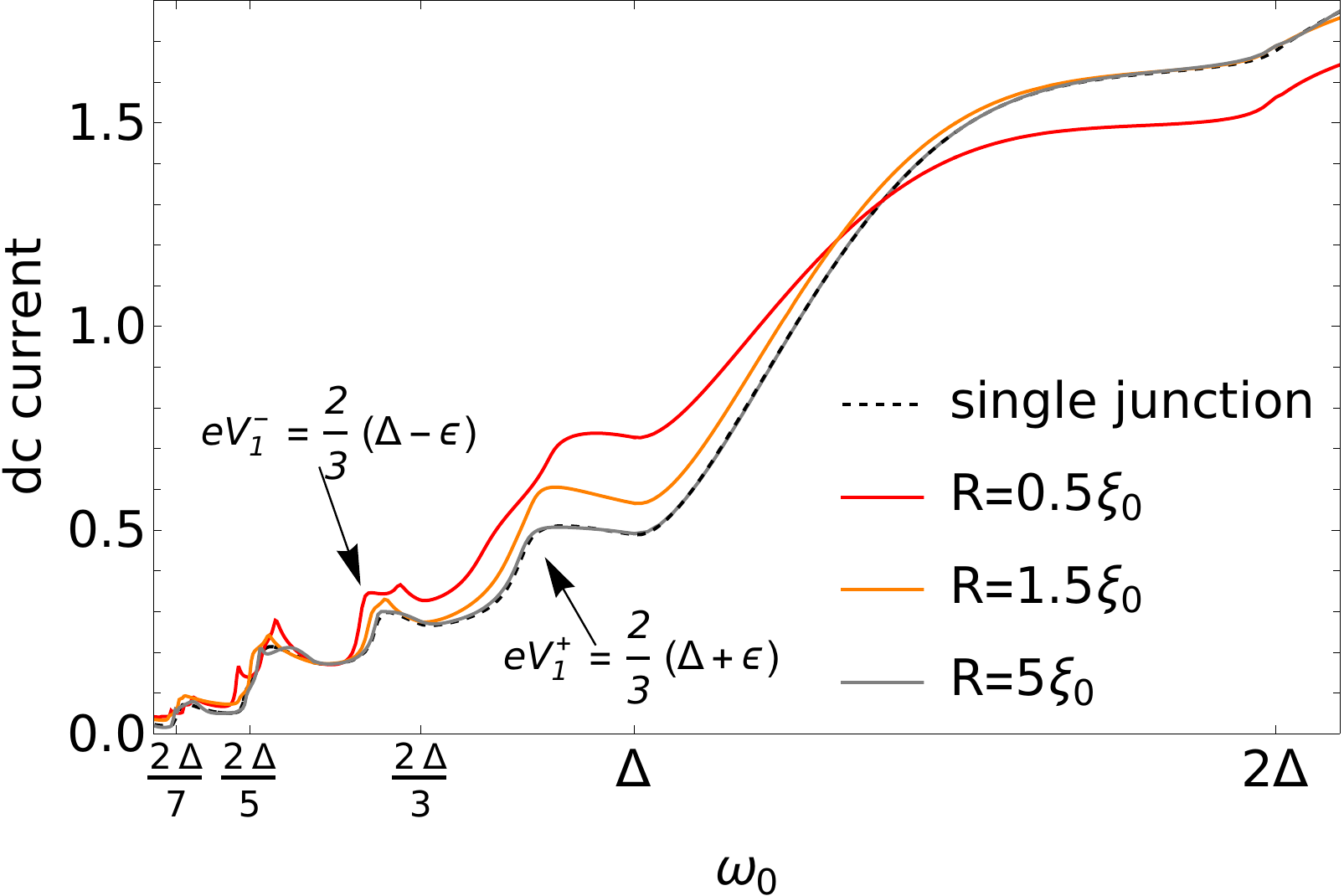}
	\caption{Andreev molecule I--V curve calculated at various distances between the two dots. Parameters used: $\Gamma_{a,b,c}=0.3\Delta, k_FR=\pi/4, \phi_{a,b,c}=0.$}
	\label{twodotscurrent}
\end{figure}
\begin{figure}
	\centering
	\includegraphics[width=0.9\linewidth]{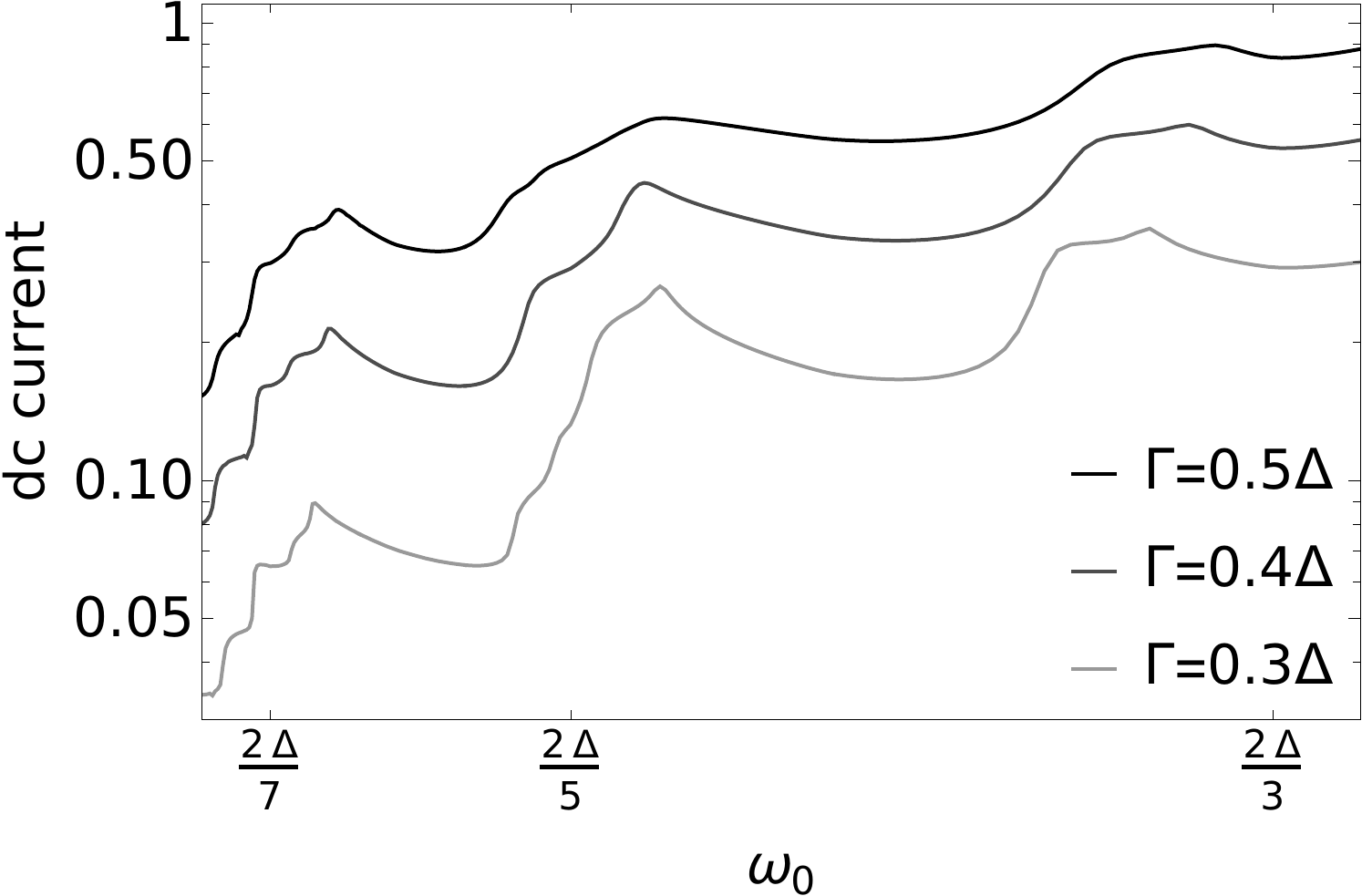}
	\caption{Andreev molecule I--V curves for different values of tunnel coupling $\Gamma_{a,b,c}=\Gamma,$ at fixed distance between the dots $R=\xi_0.$ Logarithmic scaling is used. Features are softened with increasing $\Gamma$ and voltage.}
	\label{couplings}
\end{figure}
\begin{figure}
	\centering
	\begin{subfigure}{0.8\linewidth}
		\centering
		\includegraphics[width=\textwidth]{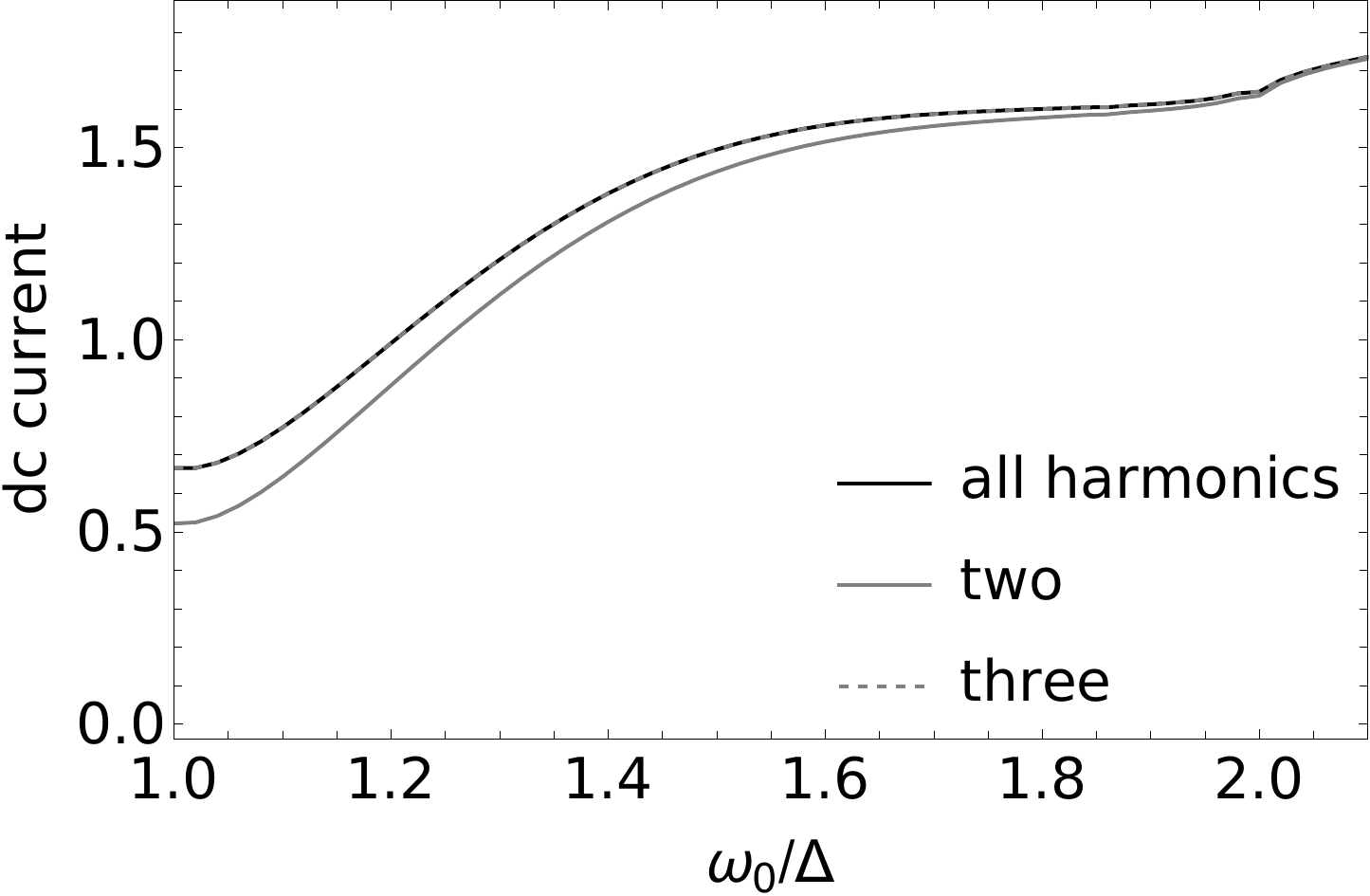}
		\caption{}
	\end{subfigure}
		\hfill
	\begin{subfigure}{0.8\linewidth}
		\centering
		\includegraphics[width=\textwidth]{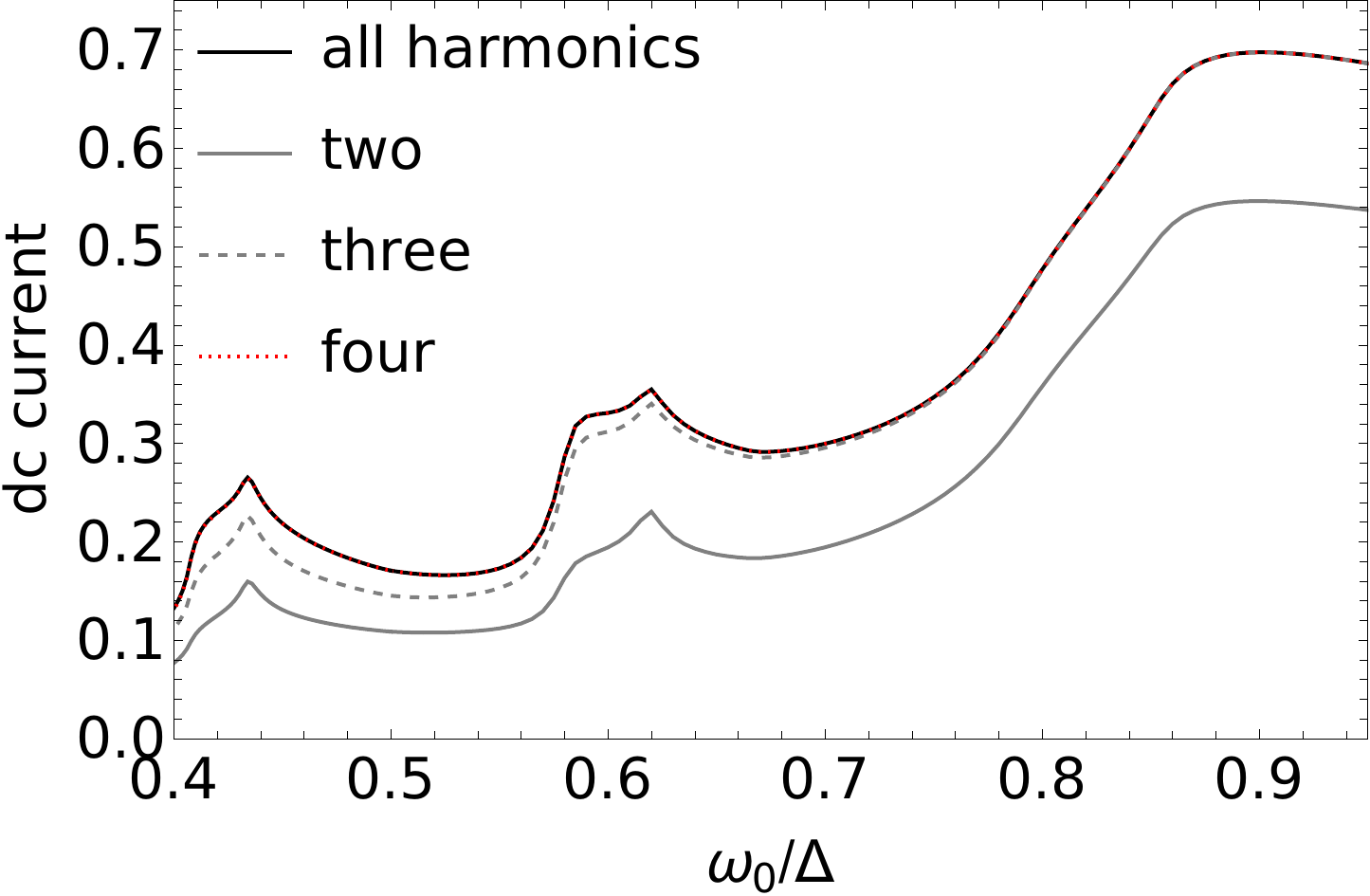}
		\caption{}
	\end{subfigure}
		\hfill
	\begin{subfigure}{0.8\linewidth}
	\centering
	\includegraphics[width=\textwidth]{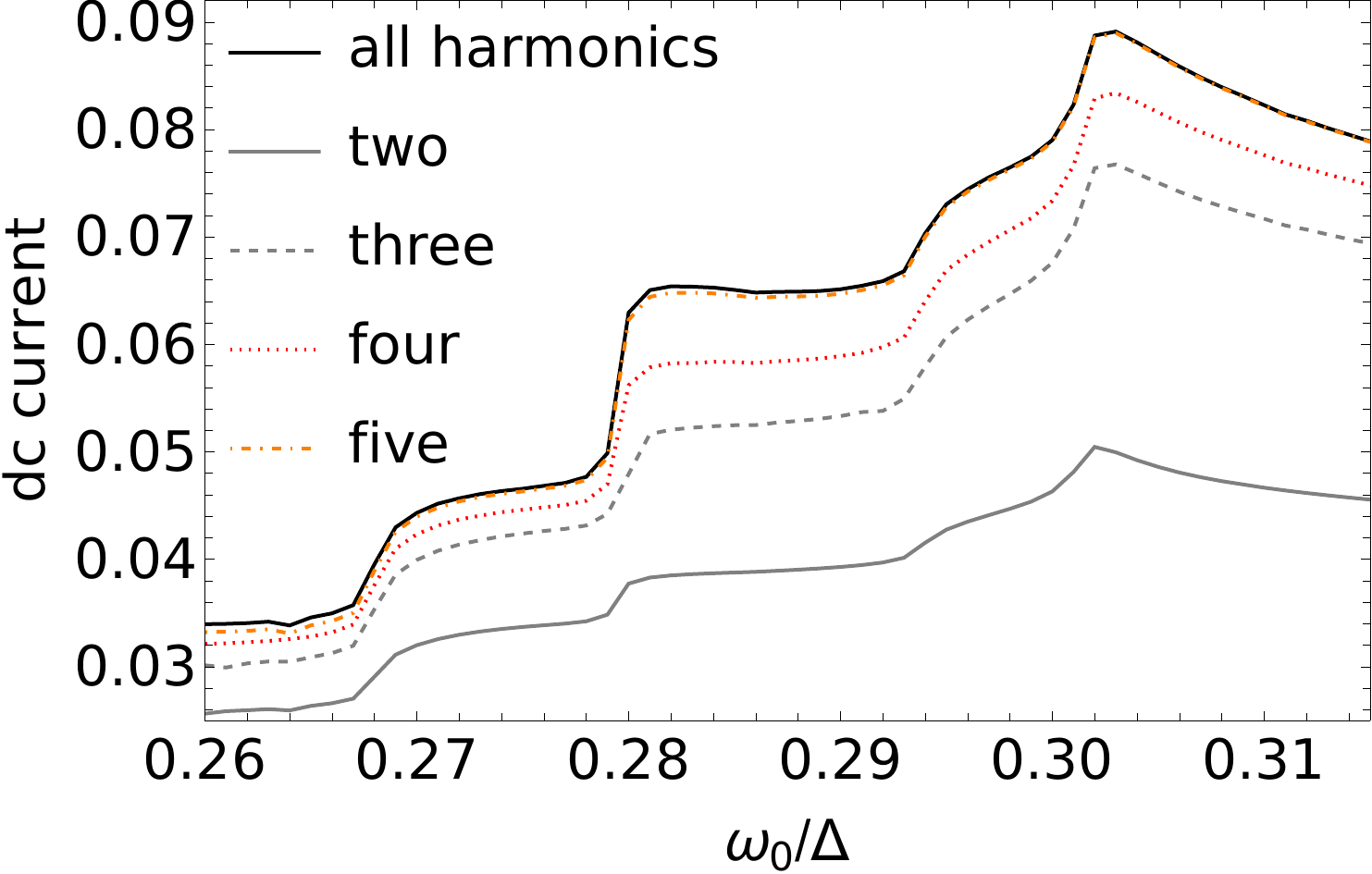}
	\caption{}
\end{subfigure}
	\caption{`Localization' on the Floquet chain means only a few of the harmonics need to be taken into account when calculating the current. The amount of harmonics required increases with decreasing voltage. Parameters: $R=\xi_0$, $\Gamma_{a,b,c}=0.3\Delta$, $k_FR=\pi/4.$}
	\label{harmonics}
\end{figure}
\paragraph{Modification of the subgap structure.} The main result for the bijunction current is presented in Fig. \ref{twodotscurrent}, where we plot the results of numerical calculations of Eq. (\ref{twodoteq}). For simplicity, we consider equal tunnel couplings $\Gamma_a=\Gamma_b=\Gamma_c=0.3\Delta,$ and we fix $k_FR=\pi/4,$ in order to avoid oscillations on the scale of the Fermi wavelength. We see that for large distances between the dots (grey line) the I--V curve approaches that of a single S-dot-S junction (dashed black line). In reality, at large distances, the curves show oscillations around the single junction curve; this will be commented on shortly. For distances comparable to the superconducting length (red and orange line), we are clearly in a `molecular junction' regime \cite{molecularjunction}, and the MAR steps break into four sub-steps. These sub-steps correspond to the splitting of the energy levels in the Floquet spectrum in the Andreev molecule regime, as discussed earlier (see Fig. \ref{bijunction spectrum}). The steps are visible when the resonances are not overlapping, that is when their widths are smaller than their separation. Given that the width of a resonance coupled to a continuum of states increases when the coupling to the continuum is increased, one expects that small values of voltage bias and tunnel couplings give sharper features. Indeed, the new features due to the proximity of a second junction are more clear at voltages equal to higher-order subdivisions of the gap. However, the modification should still be visible around the $\frac{2\Delta}{3}$ or the $\frac{2\Delta}{5}$ MAR steps.  Moreover, the influence of the tunnel couplings on the I--V curves is shown on Fig. \ref{couplings}. As it is expected we observe that the subgap features are softened when increasing the tunnel coupling to the reservoirs. 
\begin{figure}
	\centering
	\begin{subfigure}{0.9\linewidth}
		\centering
		\includegraphics[width=\textwidth]{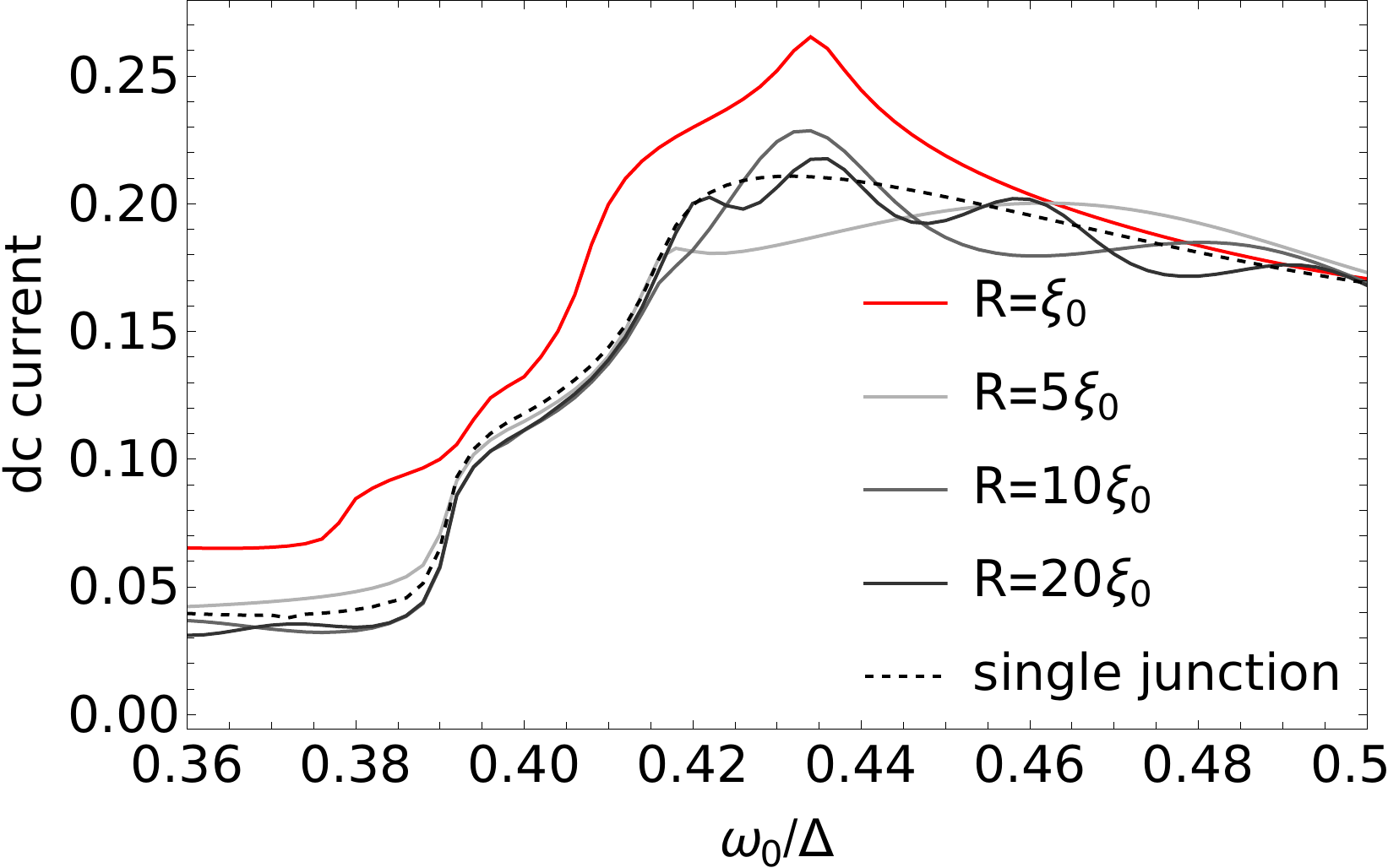}
		\caption{}
	\end{subfigure}
	\hfill
	\begin{subfigure}{0.9\linewidth}
		\centering
		\includegraphics[width=\textwidth]{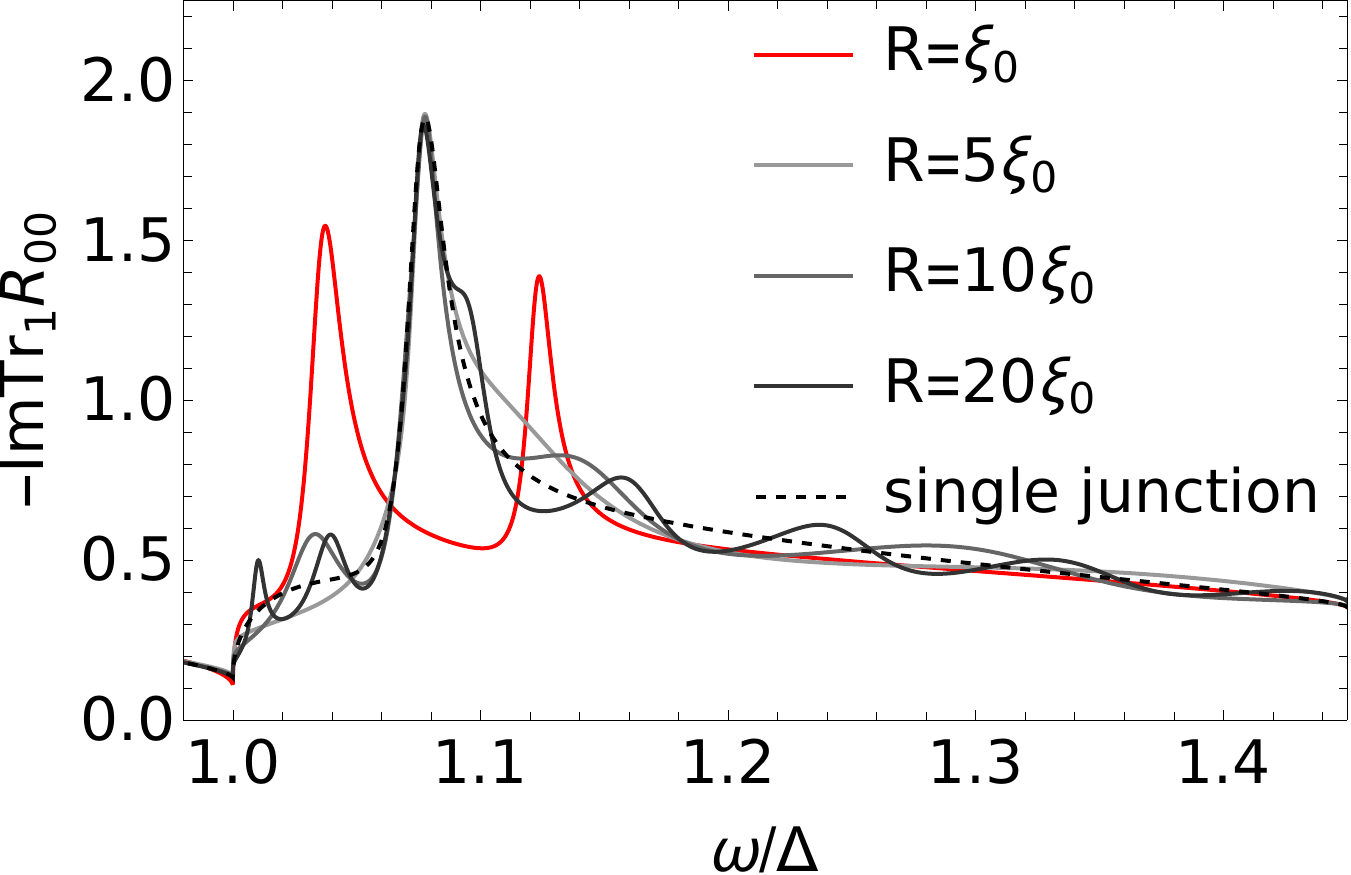}
		\caption{}
\end{subfigure}
	\caption{Long-range Floquet-Tomasch effect. (a) Subgap current structure near the second MAR step, around $\omega_0=\frac{2\Delta}{5},$ compared to the single junction case (in red). At large distances (gray lines) oscillations of the I--V curves appear around the single junction current. (b) Spectral function at energies above the superconducting gap, for $\omega_0=0.45\Delta.$}
	\label{floquet-tomasch}
\end{figure}
\paragraph{Contribution of Floquet harmonics.}
Depending on the region of the I--V curve the sum over the Floquet modes in Eq. (\ref{twodoteq}) of the current can be drastically truncated, and the larger the voltage drive, the less harmonics we need to sum over. We therefore have a `localization' on the Floquet chain, analogous to the Wannier-Stark localization of electrons in solids at strong electric fields. This localization is illustrated in Fig. \ref{harmonics}. At large voltages $\omega_0>2\Delta,$ the drive is strong enough to promote quasi-particles directly above the gap without any MAR processes, and we only need to sum over two harmonics $m=\pm 1.$ As we lower the voltage, we progressively need to add more harmonics, in correspondence to the MAR processes which are dominant. In the region of the first allowed MAR process, $\frac{2\Delta}{3}<\omega_0<2\Delta,$ the current is well approximated by summing over three harmonics $m=\pm 1, -3,$ in the next region of $\frac{2\Delta}{5}<\omega_0<\frac{2\Delta}{3},$ we need to add one more $m=\pm 1, -3, -5,$ and so on.
\subsection{Long-range Floquet-Tomasch oscillations}
The effect of the distance between the dots is shown with more detail on Fig. \ref{floquet-tomasch}, focusing in the region of the MAR step around $\omega_0=\frac{2\Delta}{5}.$ We observe that an increased distance (grey lines) produces oscillations of the I--V curve itself around the single junction curve (dashed line). Moreover, the splitting between the Floquet-Andreev resonances is suppressed exponentially with the distance (see Fig. \ref{splitting}), so that at large distances we can only see the steps corresponding to the poles of the single junction resolvent. The effect at large distances is reminiscent of the Tomasch effect. Historically, Tomasch \cite{Tomasch} observed oscillations in the density of states above the gap and in the tunneling current. The oscillations depend on the applied voltage and the thickness $d$ of superconducting films as a function of the combination $\frac{2d\sqrt{(eV)^2-\Delta^2}}{\hbar v_F}.$ Following the experimental observation, McMillan and Anderson \cite{McMillan-Anderson} then interpreted the phenomenon as one of quasi-particle interference due to a perturbation in the order parameter (induced by some impurity or by some spatially non-uniform $\Delta$). 
Figure \ref{floquet-tomasch}(b) shows the spectral function $-\Im\Tr_{\mathrm{dot 1}}\mathcal{R}_{00}$ at energies above the gap, calculated at a fixed voltage value $\omega_0=0.45\Delta.$ We observe that at large distances the spectral function oscillates around the spectral function of the single junction. We find that the frequency of oscillations \emph{is} the Tomasch frequency, and we provide a demonstration in the Appendix \ref{oscillations} to support this observation. It has been argued \cite{melin2021} that this new `Floquet-Tomasch' effect could be used to create correlations of Cooper pairs over long distances, which are orders of magnitude larger than the superconducting coherence length (in the Tomasch experiment the thickness was some tens of micrometers $10-30 \mu m\sim 100\xi_0.$). It is a non-local effect over distances which are not achievable in the absence of the voltage drive since it is mediated by quasi-particles which, by MAR processes, reach the continuum of states of the middle superconductor $S_c$ where they can propagate over a long distance without being bound by the superconducting coherence length.
\paragraph{Future work.} As our numerical results for the current suggest, the physics of the system at large dot separation is that of a Fabry-P\'{e}rot interferometer, with long-range coupling between the two dots mediated by dissipating continua. We intend to further explore this in upcoming work, particularly in relation to the spectral properties of the system.
\section{Conclusion}\label{conclusions}
In conclusion, we have studied the Andreev molecule subjected to a dc voltage drive. We find that the Floquet-Andreev spectra exhibit level-splitting at small interdot separations, while at large separations they exhibit oscillations as a function of the energy, similar to the Tomasch effect. Since transport through a dot or a molecule is closely related to its spectral properties, we look for modifications of the subgap current. At the molecular regime we find that the splitting of the Floquet-Andreev resonances corresponds to a splitting of the MAR steps into sub-steps, while at the opposite regime of large separation we find that the current is an oscillatory function of the voltage, and in this sense the system behaves as an interferometer.
This study could allow interpreting the Andreev molecule signatures in non-equilibrium dc transport experiments, an understanding that was missing so far. Moreover, the method we have employed can be adapted to accommodate various different and more complex situations, such as multi-level dots, or multi-terminal configurations.

\begin{acknowledgments}
We wish to thank \c{C}. Girit for some very helpful discussions.
B. D. also thanks R\'egis M\'elin for sharing his results on long-range Floquet-Tomasch correlations before their publication \cite{melin2021}, and for many
stimulating discussions.
\end{acknowledgments}

\appendix
\section{Oscillations above the gap}\label{oscillations}
For energies above the gap $\omega>\Delta,$ we have seen that the spectral function on dot 1, given by $$\mathcal{A}(\omega)= -2\Im\mathcal{R}_{00}^{11}(\omega)=-\Im\bqty{\mathcal{R}_{00}^{11}(\omega)+\mathcal{R}_{00}^{22}(-\omega)}$$
oscillates as a function of the energy $\omega$ and the distance $R$ separating dot 1 and dot 2, a phenomenon reminiscent of the Tomasch oscillations. In fact, we will show that the oscillations of the spectral function are due to a `Tomasch phase factor' equal to $e^{2i\sqrt{\omega^2-\Delta^2}R/v_F}$, and will explain the physical process that, in this case, gives rise to the oscillations.

In order to calculate the resolvent above the gap, we can neglect the forward self-energies $\Sigma^{+},$ given in Eq. (\ref{self-energies}).  The resolvent can then be written as 
\begin{equation}\label{resolv}
\mathcal{R}_{00}=\bqty{M^0(0)-\Sigma^{-}(0)}^{-1},
\end{equation} 
with the backward self-energy given by
\begin{equation}
\begin{split}
    \Sigma^{-}(0) &\equiv \mqty*(\Sigma^{-}_1(0) & \Sigma^{-}_{12}(0) \\ \Sigma^{-}_{21}(0) & \Sigma^{-}_2(0)) \\
    &=M^{-}(-1)\frac{1}{M^0(-2)-\Sigma^{-}(-2)}M^{+}(-1).
\end{split}
\end{equation}
We would like to compare the spectral function of the bijunction system on dot 1 to that of a single junction. We must therefore define the latter: if the distance between the dots $R\to\infty,$ then the resolvent of dot 1(2) would be:
\begin{equation}
    \mathcal{R}_{1(2)}=\bqty{M^0_{1(2)}(0)-\Sigma^{-}_{1(2)}(0)}^{-1}
\end{equation}
where all matrices are now calculated in the $2\times 2$ subspace of dot 1(2). The backward self-energies $\Sigma^{-}_{1(2)}(0)$ represent \textit{local} MAR processes which connect the state on the dot 1(2) above the gap to those below the gap.

The resolvent (\ref{resolv}) of the bijunction system can the be written as a block matrix
\begin{equation}\label{approx}
\begin{split}
    \mathcal{R}_{00} &=\mqty*(M^0_1(0)-\Sigma^{-}_1(0) & g_c(0,R)-\Sigma^{-}_{12}(0)\\ g_c(0,R)-\Sigma^{-}_{21}(0) & M^0_2(0)-\Sigma^{-}_2(0))^{-1} \\
    &\approx \mqty*(M^0_1(0)-\Sigma^{-}_1(0) & g_c(0,R)\\ g_c(0,R) & M^0_2(0)-\Sigma^{-}_2(0))^{-1}.
\end{split}
\end{equation}
We can neglect the matrices $\Sigma^{-}_{12,21}$ in the off-diagonal, since they are of higher-order in the tunnel couplings. The resolvent (\ref{approx}) therefore describes resonances on each dot which are formed due to local MAR processes and which are then coupled by propagation in the middle reservoir, represented by the non-local Green's function $g_c(0,R)$ in the off-diagonal.\\
Since the resolvent is a block matrix, we can invert it blockwise\footnote{If $M$ is the inverse of a block matrix
$$M=\mqty*(A & B \\ C & D)^{-1}=\mqty*(M_{11} & M_{12} \\ M_{21} & M_{22})$$
then its upper-left block is equal to $$M_{11}=(A-BD^{-1}C)^{-1}.$$}. For the resolvent of Eq. (\ref{approx}), the upper-left block (corresponding to the first dot) is
\begin{equation}
\begin{split}
    \bqty{\mathcal{R}_{00}}_{\mathrm{dot 1}} &=\frac{\mathcal{R}_1}{1-\mathcal{R}_1 g_c(0,R) \mathcal{R}_2 g_c(0,R)} \\
    &\approx \mathcal{R}_1 +\mathcal{R}_1 g_c(0,R)\mathcal{R}_2 g_c(0,R)\mathcal{R}_1 + \dots
\end{split}
\end{equation}
We therefore find that the first correction to the resolvent of the two coupled dots with respect to the resolvent corresponding to an uncoupled dot is
\begin{equation}
\bqty{\mathcal{R}_{00}}_{\mathrm{dot 1}}-\mathcal{R}_1 \approx  \mathcal{R}_1 g_c(0,R)\mathcal{R}_2 g_c(0,R)\mathcal{R}_1.
\end{equation}

A term like the above has a clear physical interpretation: the two resonances represented by $\mathcal{R}_{1,2}$ are coupled via propagating quasi-particles in the continuum of states of the middle superconductor. The amplitude of the effect will therefore depend on the specific geometry of the middle reservoir (here we have considered an one-dimensional superconducting wire). We see that the Tomasch phase factor will appear, since the non-local Green's function $g_c(0,R)$ is proportional to a phase $e^{i\sqrt{\omega^2-\Delta^2}R/v_F}$. A phase $e^{2i\sqrt{\omega^2-\Delta^2}R/v_F}$ is then accumulated by quasi-particles which travel from resonance $1$ to resonance $2$ and back. Therefore, at a fixed distance $R,$ the resolvent of the coupled system oscillates as a function of the energy around the single-dot resolvent $\mathcal{R}_1.$ 

\bibliography{bibliography}
\end{document}